\documentclass[DIV12,10pt]{scrartcl}
\usepackage[utf8]{inputenc}
\usepackage[english]{babel}

\usepackage{amsmath,blkarray}

\usepackage[numbers]{natbib}

\usepackage{amsthm,amssymb,amsmath}
\usepackage[noblocks]{authblk}

\usepackage{multirow}
\usepackage{multicol}
\usepackage{graphicx}

\usepackage{comment}
\usepackage{caption}
\usepackage{color}
\usepackage{setspace}

\usepackage{tikz}
\newcommand{\Cross}{\mathbin{\tikz [x=1ex,y=1ex,line width=.1ex] \draw (0,0) -- (1,1) (0,1) -- (1,0);}}%

\setkomafont{title}{\normalfont \LARGE \bfseries}
\setkomafont{section}{\normalfont \Large \bfseries \boldmath}
\setkomafont{subsection}{\normalfont \large \bfseries}
\setkomafont{subsubsection}{\normalfont \normalsize \bfseries}
\setkomafont{descriptionlabel}{\itshape}
\setkomafont{paragraph}{\normalfont \bfseries \boldmath}

\newtheorem{algorithm}{Algorithm}

\usepackage{fancyhdr}
\fancypagestyle{myplain}
{
  \fancyhf{}

  \fancyfoot[C]{\thepage}
}
\title{Sensor and Sink Placement, Scheduling and Routing Algorithms for Connected Coverage of Wireless Sensor Networks} 

\author[1]{Banu Kabakulak\thanks{Corresponding author. Tel.: +90 2123596771; fax: +90 2122651800. \\ E-mail addresses: banu.kabakulak@boun.edu.tr (B. Kabakulak).}}

\affil[1]{Department of Industrial Engineering, Bo\u{g}azi\c{c}i University, \.{I}stanbul, Turkey}

\date{\vspace*{-2em}}

\begin{document}

\maketitle

\thispagestyle{empty}

\begin{abstract}

\vspace{-2mm}

A sensor is a small electronic device which has the ability to sense, compute and communicate either with other sensors or directly with a base station (sink).
In a wireless sensor network (WSN), the sensors monitor a region and transmit the collected data packets through routes to the sinks. In this study, we propose a mixed--integer linear programming (MILP) model to maximize the number of time periods that a WSN carries out the desired tasks with limited energy and budget. Our sink and sensor placement, scheduling, routing with connected coverage ($SPSRC$) model is the first in the literature that combines the decisions for the locations of sinks and sensors, activity schedules of the deployed sensors, and data flow routes from each active sensor to its assigned sink for connected coverage of the network over a finite planning horizon.
 The problem is NP--hard and difficult to solve even for small instances. Assuming that the sink locations are known, we develop heuristics which construct a feasible solution of the problem by gradually satisfying the constraints. Then, we introduce search heuristics to determine the locations of the sinks to maximize the network lifetime. Computational experiments reveal that our heuristic methods can find near optimal solutions in an acceptable amount of time compared to the commercial solver CPLEX 12.7.0. 


\end{abstract}

\newpage
\section{Introduction} \label{Introduction}


Wireless sensor networks (WSNs) are composed of a large number of wireless devices, called $sensor$s, equipped with communication and computing capabilities to monitor a $region$. A $homogeneous$ WSN consists of identical sensors, whereas the communication and computing capability of the sensors are different in a $heterogeneous$ network. WSNs are applied to various fields of technology thanks to their easy and cheap deployment features. They are used to gather information about human activities in health care, battlefield surveillance in military, monitor wildlife or pollution in environmental sciences, and so on \cite{YSA+14}.  


A sensor can collect data within its \textit{sensing range}, process data as packets and transmit to a base station ($sink$) either directly or through other sensors which are within its \textit{communication range}. Sensors consume energy for $sensing$, $receiving$ data from other sensors and $transmitting$ data to other sensors or a sink. Energy-aware operating is important for a sensor since it has limited battery energy. A sensor can carry out sensing and communicating tasks when it is $active$ and consumes negligible energy in $standby$ mode \cite{LSS10}. A sensor is no more a member of the WSN, when its battery energy depletes.


The number of time periods that a WSN operates as desired is its $lifetime$ and depends highly on the limited energy of the sensors. Hence, energy--aware usage of the sensors helps to prolong the network lifetime. The key factors that affect the energy consumption can be listed as follows: locations of the sensors and sinks in the network, schedule of the active or standby periods of the sensors, sink assignments of the sensors and data transmission routes from the sensors to their assigned sinks. 

Design problems related with WSNs are well studied in the literature. The Coverage Problem (CP) focuses on how well the sensors observe the region they deployed. Each node in the region has a \emph{coverage requirement}, i.e., least number of sensors that should monitor the node. In order to minimize the energy consumption, CP aims to locate smallest number of sensors in the region to satisfy the coverage requirements of the nodes \cite{YHL+15, ZZS+12}. The reviews \cite{GKJ16, GY16} summarize the works on CP. A genetic algorithm is proposed in \cite{TGG16} to cover each node with at least one sensor. The greedy algorithm in \cite{CDR12} provides coverage when the sensing ranges of the sensors are adjustable. In \cite{LRS18}, the column generation algorithm locates the sensors while minimizing the sensing energy consumption of the sensors.  In \cite{TGG14}, mobile robots are used in cooperation with a WSN to monitor a disaster area. In this study, we consider \emph{differentiated} CP, where each node in the network has different coverage requirement. 

The active sensors should be connected to a sink to transmit the sensed data.  Coverage and connectivity are important for a WSN, since they are quality of service (QoS) measures. There are different definitions in the literature for network connectivity. A network can be considered to be connected if the active sensors can communicate at least one of the other active sensors \cite{WX06}. The authors restore the connectivity among the sensors with a heuristic algorithm in the case of a failure of sensors in \cite{LYL15}. Another definition is the existance of a path from each active sensor to \emph{a sink} in order to transmit the data packets \cite{AIH10}. Hence, the locations of the sensors are determined to satisfy the coverage requirements of the nodes and provide the connectivity in the Connected Coverage Problem (CCP). In \cite{HLJ+17}, the authors give energy efficient algorithms for CCP. There are numerous works in the literature on CCP such as \cite{ACL+10,  YTB+14, ZD11}. 
In our study, we make a broader definition of connectivity and we assume a network is $\alpha-connected$ in a period if each active sensor can communicate with $\alpha-$many active sensors and there is a data flow path to  \emph{its assigned sink}. Hence, each active sensor can communicate with its assigned sink either directly or via other active sensors by multi-hop transmission. Moreover, one can find alternative paths from the sensors to the sinks in the case of unexpected failure of a sensor in the network, since there are at least $(\alpha- 1)$ active sensors in the communication range of each sensor.

One can use the limited energy of the sensors more economically by keeping unnecessary sensors in standby mode in a period. That is, the Activity Scheduling Problem (ASP) determines the active and standby periods of the sensors in order to extend the network lifetime \cite{ RBC14, SM11}. The authors schedule the sensors in homogeneous and heterogeneous WSNs in \cite{WLL+10}.  Heuristic algorithms are developed in \cite{MUS14} to deploy sensors for CP and schedule their activities to maximize the lifetime. 
In some WSNs, such as cognitive WSNs, the sensed data packets can be processed by only some of the sink nodes.  The Sink Assignment Problem (SAP) assigns the active sensors to an appropriate sink. SAP allows to control the data traffic load on the sinks. This helps to avoid congestion, i.e., traffic load exceeds the available capacity, on the sinks as noted in \cite{KDO+14}. The authors formulate the SAP for cognitive WSNs as MILP in  \cite{WHL+17} and develop a linear relaxation technique for its solution. 

Sensors also consume energy while transmitting the sensed data to the sinks. Hence, the Data Routing Problem (DRP) finds the minimum energy consuming paths from the active sensors to the sinks \cite{ZLZ+14}. A column generation algorithm is developed in \cite{TAA+10} for CCP, ASP, and DRP.  An integer programming formulation for DRP, when the sensor locations are known, is given in \cite{YCV15}. The authors introduce centralized heuristics to minimize the energy consumption.  
In the literature, it is sufficient to find a data flow path from a sensor to ``a sink." On the other hand, we extend DRP in this study to design more secure networks. That is, we make sink assignment for each active sensor and construct a data flow path from a sensor to ``its assigned sink." Such a network design securely collects data, since only the desired sink can receive the data packets. 
The locations of the sinks also affect the energy consumption in data routing. The Sink Location Problem (SLP) investigates the optimum locations of the sinks that require the least transmission energy.  In \cite{SCN14, TID+15}, mobile sinks are considered to collect data from the sensors. The paths of the mobile sensors are also addressed in \cite{K17}. The authors develop heuristics to find the locations of the sinks at each period.  
In our study, we assume that sink nodes are immobile. 



In this study, we consider heterogeneous WSNs. We consider to deploy sensor in the network to satisfy differentiated coverage requirements of the nodes and $\alpha-$connectivity of the sensors in CCP. We schedule the active and standby periods of the sensors for energy--aware design of the network in ASP. We assign a sink for each active sensor in a period in SAP. We detect minimum energy consuming paths from the active sensors to their assigned sinks in DRP. We find the best locations of the sinks to maximize the network lifetime in SLP. We name the network design problem that combines CCP, ASP, SAP, and DRP as the Sensor Placement, Scheduling and Routing with Connectivity ($PSRC$) problem. Additionally, we have the $SPSRC$ problem, which deals with SLP and $PSRC$ simultaneously. $SPSRC$ is difficult to solve exactly for large networks since it includes the set covering problem, which is known to be NP--complete \cite{GJ90}, as a subproblem. 

We can list our contribution to the literature as follows:

\begin{itemize}
	\item We extend the definitions of CCP and DRP by introducing $\alpha-$connectivity and SAP.
	\item This paper makes original contributions to the literature by dealing with $SPSRC$ model which integrates SLP, CCP, ASP, SAP, and DRP. To the best of our knowledge, the design issues mentioned are not undertaken within a single model in the literature before.

	\item We reduce $SPSRC$ model to $PSRC$ by assuming that the sink locations are known. We propose Constructive Heuristic (CH) and Disjunctive Heuristic (DH) methods for the solution of $PSRC$.

	\item We generate feasible solutions of $SPSRC$ with our Local Search (LS) and Tabu Search (TS) algorithms, which determine the sink locations. 
  
	\item  Our solution methods can provide near optimal feasible solutions of $SPSRC$ in an acceptable amount of time.  Our algorithms outperform CPLEX 12.7.0 in terms of solution quality and computation time. 

\end{itemize}

In the remainder of the paper, we  formally define our problem in Section \ref{ProblemDefinition} and introduce our $SPSRC$ formulation in Section \ref{sec:MathematicalFormulations}. We give the details of our heuristic methods for the solution of $SPSRC$ in Section \ref{SolutionMethods}. We test the efficiacy of our methods via computational experiments in Section \ref{ComputationalResults}. Some concluding remarks and comments on future work appear in Section \ref{Conclusions}.  

\section{Problem Definition}\label{ProblemDefinition}

In this paper, we consider a heterogeneous WSN with $K$ sensor types given with set $\mathbf{K} = \{1, ..., K\}$. We assume sink is type--0 and set $\mathbf{\bar{K}} = \mathbf{K}\bigcup \{0\}$ represents the sensor and sink types. WSN monitors a region defined by a set of nodes, i.e., $\mathbf{N} = \{1, ..., N \}$. We deploy and activate sensors at each time period $t \in \mathbf{T} = \{1, ..., T \}$ to cover each node $i \in \mathbf{N}$ with at least $f_i-$many sensors. The cost of placing a sensor $(j, k)$, i.e., type-$k$ sensor at location $j \in \mathbf{N}$, is given as $c_{jk}$. We have a total budget of $B$ monetary units. 

An active sensor $(j, k)$ can cover nodes within its sensing ($r_k^s$) range and generates $h_{jk}-$many data packets in a period. The coverage coefficient $a_{jki}$ is one if sensor $(j, k)$ can cover node $i$, and zero otherwise.  
A type-$k$ sensor can receive and transmit data packets within its communication ($r_k^c$) range. The communication coefficient $b_{jki}$ is one if sensor $(j, k)$ can communicate with a sensor or a sink at node $i$, and zero otherwise. In order to have a robust communication, we require that an active sensor communicates with at least $\alpha-$many active sensors in a period. A type-$k$ sensor has $E_k$ units of energy initially. It consumes $e_k^s$ units for sensing the network, $e_k^r$ units for receiving and $e_k^c$ units for transmitting data packets in its active periods. 

At each period $t$ within network lifetime $L$, we aim to cover each node $i$ and route the data packets collected by the active sensors to their assigned sinks. We consider $T$ as the planning horizon, which is an upper bound on the network lifetime $L$.
We assume that a sink receives data packets but cannot transmit them to other sensors or sinks. That is when there is a sink $(j, 0)$, we have $b_{j0i}$ equal to zero for all $i \in \mathbf{N}$. Furthermore, sinks cannot contribute to the coverage of nodes. Hence, for a sink $(j, 0)$, we have $a_{j0i}$ equal to zero for all $i \in \mathbf{N}$. 

\begin{figure}[!h]
\begin{center}
	\includegraphics[width=0.95\columnwidth]{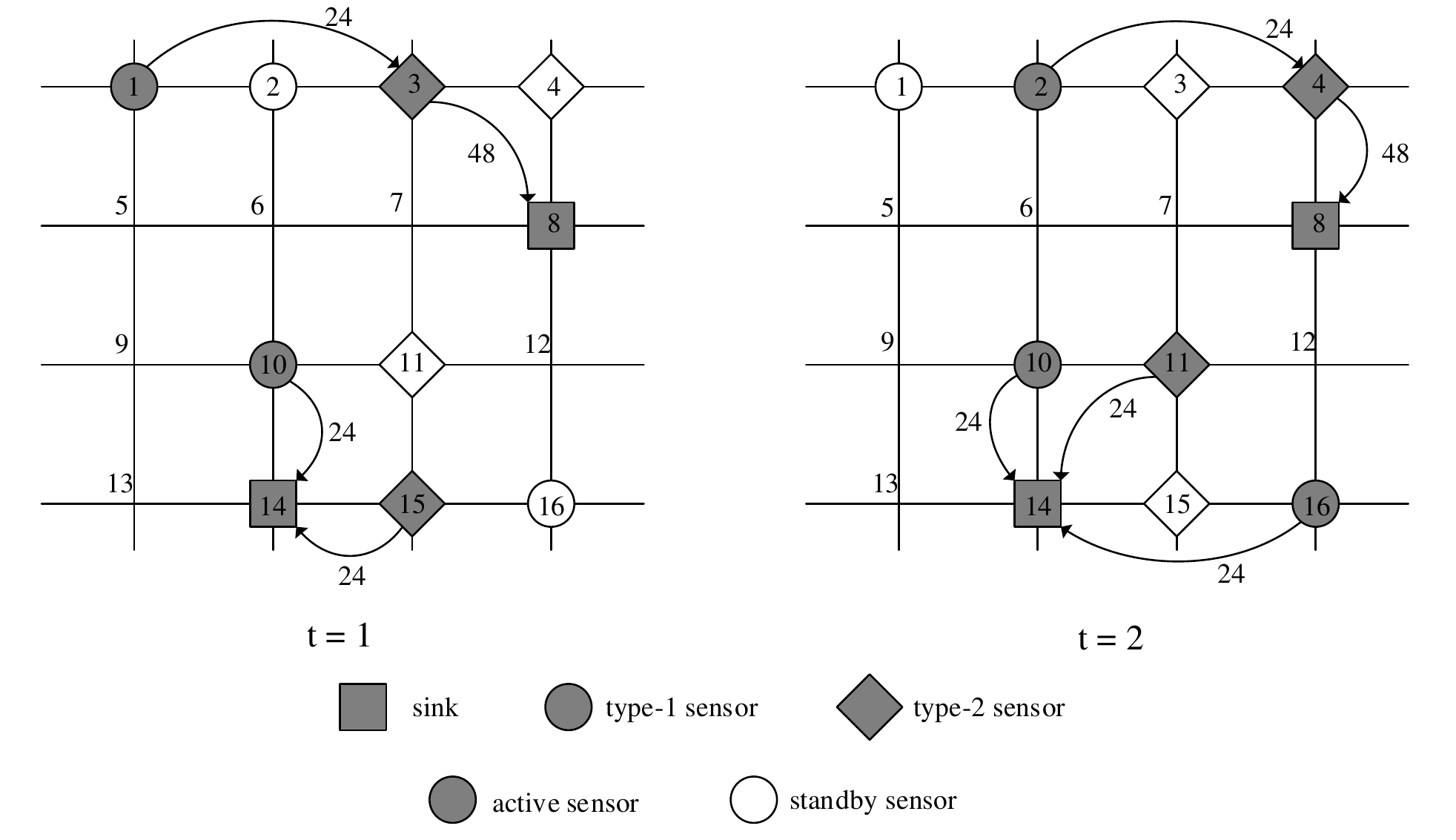}
\end{center}
\caption{A sample sensor network operating for $T = 2$}
\label{fig:network}
\end{figure}

We illustrate a feasible solution of the $SPSRC$ problem with planning horizon $T = 2$ in Figure \ref{fig:network}. In this example, we have a $4 \times 4$ grid network, i.e., $N = 16$ candidate nodes to deploy sinks and sensors. We have two different types of sensors ($K = 2$) and two sinks to collect the data. Coverage requirement is one ($f_i = 1$) for each node $i \in \mathbf{N}$. The sensing range of type--1 sensors is one, and half of the sensing range of type--2 sensors, i.e., $2r_1^s = r_2^s = 2$. The communication range of type--1 sensors is two, and half of the communication range of type--2 sensors, i.e., $2r_1^c = r_2^c = 4$.  Each active sensor communicates with one other active sensor ($\alpha = 1$) and generates $h_{jk} = 24$ data packets in a period.

In the feasible solution given in Figure \ref{fig:network}, the decisions for the network are as follows:  

\begin{itemize}
	\item (SLP) Two sinks are located at nodes 8 and 14  (rectangles), i.e., $(8, 0)$ and $(14, 0)$. They are active throughout the planning horizon $T$. 
	\item (CCP) We deploy four type--1 sensors (circles), i.e., $(1, 1), (2, 1), (10, 1)$ and $(16, 1)$. There are four type--2 sensors (diamonds), which are $(3, 2),( 4, 2), (11, 2)$ and $(15, 2)$. 
	\item (ASP) In period $t = 1$, the sensors $(1, 1), (10, 1), (3, 2)$ and $(15, 2)$ are active (black). The sensors $(2, 1), (16, 1), (4, 2)$ and $(11, 2)$ are in standby mode (white) at $t = 1$.  In period $t = 2$, the sensors  $(2, 1), (10, 1), (16, 1), (4, 2),$ and $(11, 2)$ are active, whereas $(1, 1), (3, 2)$ and $(15, 2)$ are in standby mode. 

One can observe that each node is within the sensing range of at least one active sensor ($f_i = 1$). Furthermore, there is at least one active sensor ($\alpha = 1$) in the communication range of each active sensor at each period.

	\item (SAP) The active sensors $(1, 1)$ and $(3, 2)$ are assigned to the sink $(8, 0)$ at $t = 1$. Besides, $(10, 1)$ and $(15, 2)$ are assigned to the sink $(14, 0)$ in the same period. At $t = 2$, the active sensors $(2, 1)$ and $(4, 2)$ transmit to the sink $(8, 0)$, and the sink  $(14, 0)$ collects data from the sensors $(10, 1), (16, 1)$ and $(11, 2)$. 
	\item (DRP) In period $t = 1$,  the sensor $(1, 1)$ transmits 24 data packets to its sink $(8, 0)$ at 2--hops. The other active sensors can send data to their sinks directly. In period $t = 2$, the sensor $(2, 1)$ connects with its sink $(8, 0)$ via sensor $(4, 2)$, whereas the other sensors can communicate with their sinks at 1--hop. 

The solution of DRP provides the paths that connect the sensors to their assigned sinks.
\end{itemize}

\vspace{-5mm}
\section{Mathematical Formulations} \label{sec:MathematicalFormulations}

In this section we introduce a MILP model, i.e, $SPSRC$, that locates sinks and sensors, determine activity schedules of the sensors, assigns a sink for each active sensor and determines the sensor-to-sink data flow routes while maximizing the network lifetime. In the case of  known sink locations, $SPSRC$ reduces to the sensor placement, scheduling and routing with connected coverage, i.e., $PSRC$, model. 




We have eight decision variables in our mathematical model. The continuous variable $L$ is the network lifetime that we aim to maximize. The binary decision variable $n_t$ is one if a period $t$ is within the lifetime $L$, and zero otherwise. The variables $x_{jk}$, $z_{jkt}$ and $u_{ijkt}$ are binary. The variable $x_{jk}$ is one if there is a type-$k$ sensor at location $j$, $z_{jkt}$ is one if the deployed sensor $(j, k)$ is active in period $t$, and $u_{ijkt}$ is one if the active sensor $(j, k)$ is assigned to the sink at location $i$ in period $t$. The amount of flow from sensor $(i, l)$ to sensor $(j, k)$ in period $t$ is given with the continuous variable $y_{iljkt}$. The continuous variable $g_{iljt}$ represents the incoming flow to a sink $(j, 0)$ from sensor $(i, l)$ in period $t$. The binary variable $w_{iljkt}$ is one if there is a positive flow from sensor $(i, l)$ to sensor/sink $(j, k)$ in period $t$, and zero otherwise.

We assume that a sink has sufficiently large battery energy. Hence, a sink is active during the network lifetime. 
 We summarize the terminology used in this paper in Table \ref{tab:lop}.

\begin{table}[t]
\captionof{table}{List of symbols}
    \label{tab:lop}
\line(1,0){460}
			\flushleft
			\vspace{-0.2in}
\begin{onehalfspace}
\footnotesize
			\begin{multicols}{2}
				\begin{tabular}[t]{l p{2.5in}}
		\multicolumn{2}{l}{\textit{Index sets}} \\
		 & \\ 
		$\mathbf{N}$ & set of sensor and sink locations \\
		$\mathbf{K}$ & set of sensor types \\
		$\mathbf{\bar{K}}$ &  set of sensor and sink types \\
		$\mathbf{T}$ &  set of periods \\
		 & \\ 
				\multicolumn{2}{l}{\textit{Parameters}} \\
				 &  \\
				$a_{ijk}$ & 1 if node $i$ is within the coverage range of sensor $(j, k)$, 0 otherwise  \\
				$b_{ilj}$ & 1 if a sensor located at node $j$ is within the communication range of sensor $(i, l)$, 0 otherwise  \\ 
				$B$ & total available budget  \\
				$c_{jk}$ & cost of placing a type-$k$ sensor at node $j$  \\ 
				$e_{k}^{c}$ & energy consumption of a type-$k$ sensor for transmitting one unit of flow  \\
				$e_{k}^{r}$ & energy consumption of a type-$k$ sensor for receiving one unit of flow  \\
				$e_{k}^{s}$ & energy consumption of a type-$k$ sensor for sensing and processing during a period \\
				$E_{k}$ & initial battery energy of a type-$k$ sensor \\
				$\alpha$ & minimum number of sensors that a sensor can communicate 
			\end{tabular}
	
			\textit{ }
			\begin{tabular}[t]{l p{2.5in}}
				$f_{i}$ & coverage quality requirement for node $i$ \\
				$h_{jk}$ & number of data packets generated by sensor $(j, k)$ per period  \\ 
				$N$ & number of candidate locations for sensors and sinks  \\
				$K$ & number of sensor types  \\
				$r_{k}^{c}$ & communication range of a type-$k$ sensor \\
				$r_{k}^{s}$ & sensing range of a type-$k$ sensor \\
				$S$ & number of sinks \\
				$T$ & planning horizon \\
				 &  \\
				\multicolumn{2}{l}{\textit{Decision Variables}}  \\
				 & \\
				$L$ & lifetime of the WSN \\
				$n_{t}$ & 1 if period $t$ is within the lifetime $L$, 0 otherwise \\
				$x_{jk}$ & 1 if a type-$k$ sensor is placed at node $j$, 0 otherwise  \\
				$z_{jkt}$ & 1 if a sensor $(j, k)$ is active in period $t$, 0 otherwise \\ 
				$u_{ijkt}$ & 1 if a sensor $(j, k)$ is assigned to a sink located at node $i$ in period $t$, 0 otherwise \\
				$y_{iljkt}$ & amount of data flow from sensor $(i, l)$ to sensor $(j, k)$ in period $t$ \\
				$g_{iljt}$ & amount of data flow from sensor $(i, l)$ to sink $(j, 0)$ in period $t$ \\
				$w_{iljkt}$ & 1 if data flows from sensor $(i, l)$ to sensor $(j, k)$ in period $t$, 0 otherwise \\
				 & 
			\end{tabular}			
	\end{multicols}	
\end{onehalfspace}
\line(1,0){460}
\end{table}

One can observe that setting all decision variables to zero gives a trivial solution with lifetime $L= 0$. Constraint (\ref{bound}) eliminates this solution from the solution space. We have $n_t = 0$ when $t > L$,  which implies $L \leq t- 1$ in constraints (\ref{period}). 

\vspace{-10mm}
\begin{eqnarray}
& 1 \leq   L  \leq T & \  \label{bound} \\
& T n_t  \geq L + 1 - t, \;\;  &  t \in \mathbf{T} \label{period}
\end{eqnarray}

At each period $t$ within lifetime $L$, we should cover a point $i \in \mathbf{N}$ with at least $f_i-$many active sensors as given in constraints (\ref{coverage}).   

\vspace{-5mm}
\begin{equation}
\sum_{(j, k) \in \mathbf{N} \Cross \mathbf{K}} a_{ijk}z_{jkt} \geq f_{i}n_{t}, \;\;  i  \in \mathbf{N},\ t\in \mathbf{T} \label{coverage} 
\end{equation}

Constraints (\ref{feasiblez}) force to deploy a sensor before activating it and constraints (\ref{feasz}) provide not to activate a sensor for the periods out of the lifetime $L$. 

\vspace{-10mm}
\begin{eqnarray}
z_{jkt} \leq x_{jk}, \;\; & (j, k) \in \mathbf{N} \Cross \mathbf{K}, \ t\in \mathbf{T} \label{feasiblez} \\
z_{jkt} \leq n_t, \;\; & (j, k) \in \mathbf{N} \Cross \mathbf{K}, \ t\in \mathbf{T} \label{feasz} 
\end{eqnarray}

In order to have a $\alpha-$connected network, as given in constraints (\ref{connectivity}), there should be at least $\alpha$--many active sensors that an active sensor $(i, l)$ can communicate at a period $t$. 

\begin{equation}
\underset{\begin{subarray}{c}
  (j, k) \in \mathbf{N} \Cross \mathbf{K} \\
  (j, k) \neq (i, l)
  \end{subarray}}{\sum} b_{ilj}z_{jkt} \geq \alpha z_{ilt},  \;\; (i, l)  \in  \mathbf{N} \Cross \mathbf{K}, t \in \mathbf{T} \label{connectivity}  
\end{equation}

We make the unique sink assignments of the active sensors in a period with constraints (\ref{feasibleu1})--(\ref{feasibleu3}). 
Constraints (\ref{feasibleu1}) require a sink $(i, 0)$ is deployed and constraints (\ref{feasibleu2}) guarantee that the sensor $(j, k)$ is active for assigning the sensor $(j, k)$ to the sink $(i, 0)$. Moreover, constraints (\ref{feasibleu3}) assign one and only one sink for each active sensor $(j, k)$ in a period $t$. 

\vspace{-10mm}
\begin{eqnarray}
u_{ijkt} \leq x_{i0}, \;\; &  i \in \mathbf{N}, (j, k) \in \mathbf{N} \Cross \mathbf{K}, \ t\in \mathbf{T}  \label{feasibleu1} \\
u_{ijkt} \leq z_{jkt}, \;\; &  i \in \mathbf{N}, (j, k) \in \mathbf{N} \Cross \mathbf{K}, \ t\in \mathbf{T} \label{feasibleu2} \\
\sum_{i\in \mathbf{N}} u_{ijkt} = z_{jkt}, \;\; & (j, k) \in \mathbf{N} \Cross \mathbf{K}, \ t\in \mathbf{T}  \label{feasibleu3} 
\end{eqnarray}

We avoid flows from a sensor $(i, l)$ to itself with constraints (\ref{feasibleflow1}). We assume that there can be only one active sensor of type-$k$ at a node in a period, since most of the time an energy efficient optimal solution has such an activity schedule. Hence, the maximum number of sensors that can be active in a period is given by $NK$. Then, an active sensor $(j, k)$ can send the total flow of $(NK - 1)$ sensors to a sensor $(i, l)$. From here, we can bound the total outflow from an active sensor $(j, k)$ by $M_1 = \underset{\begin{subarray}{c}
(j, k) \in \mathbf{N} \Cross \mathbf{K}
  \end{subarray}}{\max}{\left\{h_{jk}\right\}} (NK - 1)$.
A sensor $(i, l)$ can send flow to the sensors that are within its communication range as given in constraints (\ref{feasibleflow}). Constraints (\ref{flowM1}) guarantee that there is no outflow from a sensor $(j, k)$ in standby mode.  Similarly, total inflow to an active sensor $(j, k)$  can be at most $M_1$ as given in constraints (\ref{flowM}). 

\vspace{-10mm}
\begin{eqnarray}
y_{ililt} = 0, \;\; &  (i, l) \in \mathbf{N} \Cross \mathbf{K}, \ t\in \mathbf{T} \label{feasibleflow1} \\
y_{iljkt} \leq M_1b_{ilj}, \;\; &  (i, l),  (j, k) \in \mathbf{N} \Cross \mathbf{K}, \ t\in \mathbf{T} \label{feasibleflow} \\
\sum_{(i, l)  \in  \mathbf{N} \Cross \mathbf{K}} y_{jkilt} \leq M_1 z_{jkt}, \;\; & (j, k) \in \mathbf{N} \Cross \mathbf{K}, \ t\in \mathbf{T} \label{flowM1} \\ 
\sum_{(i, l)  \in  \mathbf{N} \Cross \mathbf{K}} y_{iljkt} \leq M_1 z_{jkt}, \;\; & (j, k) \in \mathbf{N} \Cross \mathbf{K}, \ t\in \mathbf{T} \label{flowM}
\end{eqnarray}

A sink $(j, 0)$ can collect the data packets of all sensors. A sink $(j, 0)$ can receive flows from the sensors that can communicate with the sink as enforced by constraints (\ref{feasiblesinkflow}). Constraints (\ref{flowM2}) bound the total inflow to a sink $(j, 0)$ by $M_2 =\underset{\begin{subarray}{c}
(j, k) \in \mathbf{N} \Cross \mathbf{K}
  \end{subarray}}{\max}{\left\{h_{jk}\right\}} NK$.

\vspace{-10mm}
\begin{eqnarray}
g_{iljt} \leq M_2 b_{ilj}, \;\; & (i, l) \in \mathbf{N} \Cross \mathbf{K}, j \in \mathbf{N}, \ t\in \mathbf{T} \label{feasiblesinkflow} \\
\sum_{(i, l) \in \mathbf{N} \Cross \mathbf{K}}  g_{iljt} \leq M_2 x_{j0},  \;\; & j  \in \mathbf{N}, \ t\in \mathbf{T} \label{flowM2} 
\end{eqnarray}

If a sensor $(j, k)$ is active in period $t$, then it generates $h_{jk}$ units of flow, i.e., data packets, to the incoming data flow from the active sensors and sends the total flow to other active sensors or a sink as outflow. Constraints (\ref{flowbalance}) ensure the data flow balance for each active sensor $(j, k)$ in each period $t$.

\begin{equation}
\sum_{(i, l)  \in  \mathbf{N} \Cross \mathbf{K}} y_{iljkt} + h_{jk}z_{jkt} = \sum_{(i, l)  \in  \mathbf{N} \Cross \mathbf{K}} y_{jkilt} +  \sum_{i \in \mathbf{N}}  g_{jkit}, \;\;  (j, k) \in \mathbf{N} \Cross \mathbf{K}, \ t\in \mathbf{T} \label{flowbalance}
\end{equation}

We assign a sink for each active sensor $(j, k)$, which initiates $h_{jk}$ units of data flow. We expect that the data flow of each active sensor reaches to its assigned sink. Constraints (\ref{sinkflow}) determine the total inflow to a sink $(i, 0)$ as the sum of data flows of the assigned sensors at period $t$. 

\begin{equation}
\sum_{(j, k)  \in  \mathbf{N} \Cross \mathbf{K}} g_{jkit} = \sum_{(j, k)  \in  \mathbf{N} \Cross \mathbf{K}}  h_{jk}u_{ijkt}, \;\;  i \in \mathbf{N}, \ t\in \mathbf{T} \label{sinkflow} 
\end{equation}

An active sensor can send its flow to its assigned sink either directly or through the other active sensors. Then, the sink assignment of a sensor $(i, l)$ can be one of the sink assignments of the sensors that it sends flow. In constraints (\ref{ifthen_1}), the sink assignment variables $u_{vilt}$ of the sensor $(i, l)$ is bounded with the ones of the sensor $(j, k)$, i.e., $u_{vjkt}$, if there is a positive flow from the sensor $(i, l)$ to the sensor $(j, k)$. We assume that a sink $(j, 0)$ is assigned to itself at all periods. In constraints (\ref{ifthen_2}),  $\textbf{1}_j(v)$ is the indicator function, which takes value one if $v = j$, and  zero otherwise. Constraints (\ref{ifthen_2}) guarantee that the sink assignment of a sensor $(i, l)$ is the sink $(j, 0)$, if there is a positive flow among them. 

\vspace{-10mm}
\begin{eqnarray}
u_{vilt} \leq u_{vjkt} \text{ if } y_{iljkt} > 0, \;\;  & v \in \mathbf{N},  (i, l),  (j, k) \in \mathbf{N} \Cross \mathbf{K}, \ t\in \mathbf{T} \label{ifthen_1} \\
u_{vilt} \leq \mathbf{1}_j(v)x_{j0}  \text{ if } g_{iljt} > 0, \;\;  & v, j \in \mathbf{N}, (i, l) \in \mathbf{N} \Cross \mathbf{K}, \ t\in \mathbf{T} \label{ifthen_2}  
\end{eqnarray}

In a period $t$, an active sensor consumes energy for sensing and processing the data collected from the region, for receiving data from the other active sensors and transmitting the data to the other active sensors or a sink. Total consumed energy of a sensor $(j, k)$ during its active periods is limited by the initial battery energy $E_{k}$, which is modeled with the energy constraints (\ref{energy}).

\begin{equation}
\sum_{t \in \textbf{\textsl{T}}} \left(e_k^s z_{jkt} + e_k^r \sum_{(i, l)  \in  \mathbf{N} \Cross \mathbf{K}} y_{iljkt} + e_k^c\sum_{(i, l)  \in  \mathbf{N} \Cross \mathbf{K}} y_{jkilt}  + e_k^c  \sum_{i \in \mathbf{N}}  g_{jkit} \right) \leq E_{k}, \;\;  (j, k) \in \mathbf{N} \Cross \mathbf{K} \label{energy} 
\end{equation}

Constraints (\ref{budget}) force that the total deployment cost of sensors and sinks does not exceed the total available budget.

\vspace{-8mm}
\begin{equation}
\sum_{(j, k)  \in  \mathbf{N} \Cross \mathbf{\bar{K}}} c_{jk}x_{jk} \leq B  \label{budget}
\end{equation}

Constraints (\ref{sinklocate}) limit the number of sinks in the network with $S$. 

\vspace{-8mm}
\begin{equation}
\sum_{i \in \mathbf{N}}  x_{i0} = S   \label{sinklocate} 
\end{equation}

Finally, constraints (\ref{nVar}) - (\ref{gVar}) are the binary and nonnegativity restrictions on the decision variables. 

\vspace{-10mm}
\begin{eqnarray}
n_t \in\left\{0, 1\right\}, \;\; & t\in \mathbf{T} \label{nVar} \\
x_{jk} \in\left\{0, 1\right\}, \;\; & (j, k)  \in  \mathbf{N} \Cross \mathbf{\bar{K}} \label{xVar} \\
z_{jkt} \in\left\{0, 1\right\}, \;\; & (j, k)  \in  \mathbf{N} \Cross \mathbf{K}, t\in \mathbf{T} \label{zVar} \\
u_{ijkt} \in\left\{0, 1\right\}, \;\; & i  \in  \mathbf{N}, (j, k)  \in  \mathbf{N} \Cross \mathbf{K}, t\in \mathbf{T} \label{uVar} \\
y_{iljkt} \geq 0, \;\; & (i, l),  (j, k) \in  \mathbf{N} \Cross \mathbf{K}, t\in \mathbf{T} \label{yVar} \\ 
g_{iljt} \geq 0, \;\; &  (i, l) \in  \mathbf{N} \Cross \mathbf{K}, j \in  \mathbf{N}, t\in \mathbf{T} \label{gVar} 
\end{eqnarray}

We formulate the sink and sensor placement, scheduling, routing while providing connected coverage ($SPSRC$) as mixed--integer linear programming problem subject to the constraints (\ref{bound}) -- (\ref{gVar}). We aim to maximize the network lifetime (\ref{objective}) under limited energy and budget resources.

\vspace{-5mm}
\begin{equation}
\mbox{max} \;\; L \label{objective}
\end{equation}

One can observe that constraints (\ref{ifthen_1}) and (\ref{ifthen_2}) are not linear. We introduce binary variables $w_{iljkt}$, which is one if there is a positive flow from a sensor $(i, l)$ to a sensor/sink $(j, k)$ in period $t$, and zero otherwise. Then, we linearize constraints (\ref{ifthen_1}) and (\ref{ifthen_2}) as follows:

\vspace{-10mm}
\begin{eqnarray}
y_{iljkt}  \leq M_1 w_{iljkt} , \;\;  & (i, l),  (j, k) \in \mathbf{N} \Cross \mathbf{K}, \ t\in \mathbf{T}  \label{linear1} \\
g_{iljt}  \leq M_2 w_{ilj0t} , \;\;  & (i, l) \in \mathbf{N} \Cross \mathbf{K}, j \in \mathbf{N}, \ t\in \mathbf{T}  \label{linear2} \\
u_{vilt} - u_{vjkt} \leq 1 - w_{iljkt}, \;\;  & v \in \mathbf{N},  (i, l),  (j, k)  \in \mathbf{N} \Cross \mathbf{K}, \ t\in \mathbf{T} \label{linear3} \\
u_{vilt} -\mathbf{1}_j(v)x_{j0} \leq  1 - w_{ilj0t}, \;\;  & v, j \in \mathbf{N}, (i, l) \in \mathbf{N} \Cross \mathbf{K}, \ t\in \mathbf{T} \label{linear4}
\end{eqnarray}

We have battery energy limitations of the sensors as constraints (\ref{energy}). This means, there can be alternative optimum solutions of $SPSRC$ model which consume less energy if we can avoid unnecessary flow loops. In constraints (\ref{connectivity}), we know that an active sensor can communicate with at least $\alpha-$many active sensors. From here, we can save energy in the network if we limit the number of outflow paths from a sensor $(i, l)$ by $\alpha$ as in constraints (\ref{alphaoutflows}).

\vspace{-10mm}
\begin{flalign}
\sum_{(j, k)  \in  \mathbf{N} \Cross \mathbf{K}} w_{iljkt} \leq \alpha, \;\; & (i, l) \in \mathbf{N} \Cross \mathbf{K}, \ t\in \mathbf{T} \label{alphaoutflows} 
\end{flalign}

$SPSRC$ combines multiple decisions related with a network in one model.  $SPSRC$ is an NP--complete problem \cite{GJ90, S96}. Hence, we propose heuristic approaches to generate a near optimal solution of $SPSRC$ in acceptable amount of time. We first consider that we are given the locations of $S$ sinks, i.e., $x_{i0}$ values are known. We name this problem as $PSRC$. We generate feasible solutions via our constructive heuristic (CH) and disjunctive heuristic (DH) approaches for $PSRC$. Then, we find feasible solutions of $SPSRC$ with our local search (LS) and tabu search (TS) methods, which determine the locations of the sinks. We explain the details of these algorithms in the next section.   


\section{Solution Methods} \label{SolutionMethods}

We introduce CH (see Section \ref{ConstructiveHeuristic}) and DH (see Section \ref{DisjunctiveHeuristic}) algorithms to generate feasible solutions of $PSRC$. We can construct a feasible solution of $PSRC$ with CH or DH either from scratch (all$-\mathbf{0}$ solution), or improve a given feasible solution, or repair an infeasible solution. These heuristics are embedded into our sink location search algorithms, i.e., LS (see Section \ref{LocalSearch}) and TS (see Section \ref{TabuSearch}), explained in Section \ref{SinkSearch} to maximize the network lifetime $L$. 

\subsection{Feasible Solution Generation Algorithms for $PSRC$} \label{PSRCwithSink}

\subsubsection{Constructive Heuristic} \label{ConstructiveHeuristic}

In Algorithm \ref{alg:Constructive}, we explain the main steps of CH. 
In the first step of CH, we obtain a feasible solution of $PSRC$ with respect to the coverage  (\ref{coverage}) and budget (\ref{budget}) constraints with Algorithm \ref{alg:FeasCoverBudget}. Then, we satisfy the connectivity (\ref{connectivity}) and sink assignment (\ref{feasibleu1}) - (\ref{feasibleu3}) constraints by assigning sinks to the active sensors using Algorithm \ref{alg:FeasSinkAssign}. We determine the amount of flows $\mathbf{y}$ and $\mathbf{g}$ from the active sensors to their assigned sinks by solving Routing Problem ($RP(t)$) at each period $t$. 

\begin{center}
\begin{minipage}{0.65\textwidth}
\begin{onehalfspace}
\footnotesize
\begin{algorithm} \textbf{(Constructive Heuristic)}\\
\begin{tabular}{ll}
\hline
\vspace{-4mm}\\
\textbf{Input:} An instance of $PSRC$, a vector $(L, \mathbf{n}, \mathbf{x}, \mathbf{z}, \mathbf{u}, \mathbf{y}, \mathbf{g}, \mathbf{w})$  \\ 
\hline
\vspace{-4mm}\\
0.  Set $L = T$.\\ 
1. Apply Algorithm \ref{alg:FeasCoverBudget} to obtain a feasible solution subject to coverage (\ref{coverage}) \\ 
\hspace{10pt} and budget (\ref{budget}) constraints. \\ 
2. \textbf{If}  $L = 0$, \textbf{Then} Stop. \\
3. \textbf{Else} \textbf{For Each} $t \leq L$ \\
4. \hspace{15pt} Apply Algorithm \ref{alg:FeasSinkAssign} to generate a feasible solution subject to connectivity (\ref{connectivity}) \\ 
 \hspace{28pt}  and sink assignment (\ref{feasibleu1}) - (\ref{feasibleu3}) constraints in period $t$. \\ 
5. \hspace{15pt} \textbf{End For Each} \\
6. \textbf{End If}\\
7. \textbf{If} $L = 0$, \textbf{Then} Stop. \\
8.  \textbf{Else} \textbf{For Each}  $t \leq L$, solve $RP(t)$ to determine the data flows $\mathbf{y}$ and $\mathbf{g}$. \\ 
9. \textbf{End If}\\
\hline
\vspace{-4mm}\\
\textbf{Output:}  A feasible solution of $PSRC$ with lifetime $L$. \\
\hline
\end{tabular}
\label{alg:Constructive}
\end{algorithm}
\end{onehalfspace}
\end{minipage}
\end{center}

In order to satisfy the constraints (\ref{coverage}) and (\ref{budget}), we implement Algorithm \ref{alg:FeasCoverBudget}. 
We estimate the maximum possible outflow from an active sensor $(j, k)$ in period $t$ as 

\vspace{-3mm}
\begin{equation}
\zeta_t = \left(\underset{\begin{subarray}{c}
(j, k) \in \mathbf{N} \Cross \mathbf{K}
  \end{subarray}}{\sum} z_{jkt}\right) \left( \underset{\begin{subarray}{c}
(j, k) \in \mathbf{N} \Cross \mathbf{K}
  \end{subarray}}{\max} h_{jk}\right), \;\;  t  \in \mathbf{T}.
\label{MaxOutflow}
\end{equation}

 At the beginning of each period $t$, we check whether an active sensor $(j, k)$ ($z_{jkt} = 1$) has sufficient battery energy for sensing and communicating tasks or not. At period $t = 1$, we have $E^{rem}_{jkt} = E_k$. A sensor $(j, k)$ can be active in period $t$ if the remaining battery energy satisfies $E^{rem}_{jkt} \geq (e_{k}^{s} + \zeta_t (e_{k}^{r} + e_{k}^{c}))$, we set $z_{jkt} = 0$ otherwise (Step 3). We update $E^{rem}_{jkt}$ for each period $t$ using Equation (\ref{CoverageEnergy}).

\vspace{-3mm}
\begin{equation}
E^{rem}_{jk,t+1} = E^{rem}_{jkt} - z_{jkt}(e_{k}^{s} + \zeta_t (e_{k}^{r} + e_{k}^{c})), \;\;  (j, k)  \in \mathbf{N} \Cross \mathbf{K}. \label{CoverageEnergy}
\end{equation}

We check whether each node $i$ in the region is covered by $f_i-$many active sensors or not in each period $t$. If each node is covered within the budget in a period $t$,  then we move to the next period. If budget constraint is not held, we consider to remove some of the deployed sensors without harming the coverage constraints at Step 7. For this purpose, we delete sensors, i.e., set $x_{jk} = 0$, that are in standby mode until the current period since they do not contribute to the coverage of the nodes in any of the periods. Deleting sensor $(j, k)$ improves remaining budget by $c_{jk}$ monetary units. We continue with the process while the remaining budget is negative or we cannot find a sensor to delete.

After this procedure, it is possible that we could not provide budget feasibility. In this case, one can consider to delete active sensors whose removal will not harm the coverage of the nodes. This strategy is used only when we are in the first period, i.e., $t = 1$, for the sake of simplicity of the heuristic. If budget still cannot be provided then we set $L= t - 1$ and stop the algorithm (Step 8). On the other hand, if there is an undercovered node in the region, then we first try to ensure coverage in the network by activating the existing sensors. Observe that activating a standby sensor does not demand budget usage.  

\begin{center}
\begin{minipage}{0.85\textwidth}
\begin{onehalfspace}
\footnotesize
\begin{algorithm} \textbf{(Satisfying Coverage (\ref{coverage}) and Budget (\ref{budget}) Constraints)}\\ 
\begin{tabular}{ll}
\hline
\vspace{-4mm}\\
\textbf{Input:} An instance of $PSRC$, a partial (in)feasible solution $(L, \mathbf{n}, \mathbf{x}, \mathbf{z})$  \\ 
\hline
\vspace{-4mm}\\
0. Set $t = 1$. \\ 
1. \textbf{For Each} $t \leq L$ \\
2. \textbf{For Each} active sensor $(j, k)$ in period $t$ ($z_{jkt} = 1$), calculate $E^{rem}_{jkt}$ \\
3. \hspace{10pt} \textbf{If} $E^{rem}_{jkt}$ is not sufficient, \textbf{Then} $z_{jkt} = 0$. \\
4. \textbf{End For Each} \\
5. \textbf{If} every node is covered, \textbf{Then} \\
6. \hspace{10pt} \textbf{If} budget is satisfied,  \textbf{Then} $t \leftarrow t + 1$. \\
7. \hspace{10pt} \textbf{Else} order standby sensors in $[0, t]$ in nonincreasing cost \\
\hspace{22pt} and remove them until budget is satisfied. \\
8. \hspace{22pt} \textbf{If} budget is violated,  \textbf{Then} $n_t = 0, L = t - 1$ and Stop.  \\ 
9. \textbf{Else} \textbf{While} there is $CEP_{jk} > 0$ with sufficient $E^{rem}_{jkt}$ for some sensor $(j, k)$ \\
 \hspace{30pt} activate ($z_{jkt} = 1$) the standby sensor $(j, k)$ with the highest $CEP_{jk}$. \\ 
10. \hspace{18pt} \textbf{End While} \\
11.  \hspace{18pt} \textbf{While} there is a sensor  $(j, k)$ with the highest $CCR_{jk} > 0$ \\ 
12.  \hspace{35pt} \textbf{If} budget is not enough for the sensor $(j, k)$, \textbf{Then} order standby sensors  \\
\hspace{48pt} in $[0, t]$ in nonincreasing cost and remove them until enough budget is obtained. \\
13.  \hspace{30pt} \textbf{If} budget is enough, \\
14. \hspace{30pt}  \textbf{Then} deploy ($x_{jk} = 1$) and activate ($z_{jkt} = 1$) the sensor $(j, k)$ with the highest $CCR_{jk}$. \\ 
15.  \hspace{30pt} \textbf{Else} $n_t = 0, L = t - 1$ and Stop. \\
16. \hspace{30pt} \textbf{End If}\\
17. \hspace{15pt} \textbf{End While} \\
18. \textbf{End If}\\
19. \textbf{End For Each} \\
\hline
\vspace{-4mm}\\
\textbf{Output:}  A solution satisfying the constraints (\ref{coverage}) and (\ref{budget}) with lifetime $L$. \\
\hline
\end{tabular}
\label{alg:FeasCoverBudget}
\end{algorithm}
\end{onehalfspace}
\end{minipage}
\end{center}

We choose the standby sensor to activate in a gereedy way by calculating the \emph{coverage energy product}, i.e., $CEP_{jk}$, for each sensor $(j, k)$ in a period $t$. Let $(\mathbf{n}, \mathbf{x}, \mathbf{z})$ be the possibly infeasible partial initial solution given as input to Algorithm \ref{alg:FeasCoverBudget}. In a period $t$, we calculate the amount of violation in constraints (\ref{coverage}) for each node $i$ with the undercoverage $U_{i}$ values as given in Equation (\ref{Ui}).

\vspace{-3mm}
\begin{equation}
U_{i} = \max \left\{\left(f_{i}n_{t} - \sum_{(j, k) \in \mathbf{N} \Cross \mathbf{K}} a_{ijk}z_{jkt}\right), 0 \right\}, \;\;  i \in \mathbf{N} \label{Ui}
\end{equation}

For a standby sensor, it will be a reason of choice if it can cover as many nodes as possible that have positive $U_{i}$. In addition, as the sensor has more remaining energy in its battery, the need for the deployment of new sensors will be less in the future, since we can activate the sensor in these periods also. Therefore, we can define $CEP_{jk}$ for a standby sensor $(j, k)$ in a period $t$ as

\vspace{-3mm}
\begin{equation}
CEP_{jk} = \frac{\left(\underset{\begin{subarray}{c}
i  \in \mathbf{N}: U_{i} > 0
  \end{subarray}}{\sum}a_{ijk}\right)x_{jk}E^{rem}_{jkt}}{c_{jk}}, \;\;   (j, k)  \in \mathbf{N} \Cross \mathbf{K} \label{CEPjk}
\end{equation}
where $E^{rem}_{jkt}$ represents the remaining energy of a sensor $(j, k)$ in period $t$.

Then, we activate the standby sensors starting from the one with the highest positive $CEP_{jk}$ value and continue until we provide coverage constraints or we do not have standby sensors with positive $CEP_{jk}$ value (Step 9). If we could not satisfy constraints (\ref{coverage}), then we calculate a \emph{coverage cost ratio}, i.e., $CCR_{jk}$, for each sensor $(j, k)$ as  

\vspace{-3mm}
\begin{equation}
CCR_{jk} = \frac{\left(\underset{\begin{subarray}{c}
i  \in \mathbf{N}: U_{i} > 0
  \end{subarray}}{\sum}a_{ijk}\right)\left(1 - x_{jk}\right)E_{k}}{c_{jk}}, \;\;  (j, k)  \in \mathbf{N} \Cross \mathbf{K} \label{CCRjk}
\end{equation}
where $E_{k}$ represents the initial battery energy of a type-$k$ sensor. We deploy ($x_{jk} = 1$) and activate ($z_{jkt} = 1$) a new sensor $(j, k)$ in period $t$ with the highest $CCR_{jk}$ value (Step 14). In the case remaining budget is not enough to deploy the selected sensor, we try to generate sufficient budget by deleting standby sensors that are not used until the current period (Step 12). If coverage is not provided within the budget, then we set $L = t - 1$ and stop the algorithm. The computational complexity of Algorithm \ref{alg:FeasCoverBudget} is given as $\mathcal{O}(\tau_{1} N^{3}KT + \tau_{1} N^{2}KT^{2})$ where $\tau_{1} = \underset{\begin{subarray}{c}
i \in \mathbf{N}
  \end{subarray}}{\max} \left\{\underset{\begin{subarray}{c}
(j, k) \in \mathbf{N} \Cross \mathbf{K}
  \end{subarray}}{\sum} a_{ijk}\right\}$ is the maximum number of sensors that can cover a node. 

\begin{center}
\begin{minipage}{0.85\textwidth}
\begin{onehalfspace}
\footnotesize
\begin{algorithm} \textbf{(Satisfying Connectivity (\ref{connectivity}) and Sink Assignment (\ref{feasibleu1}) - (\ref{feasibleu3}) Constraints)}\\ 
\begin{tabular}{ll}
\hline
\vspace{-4mm}\\
\textbf{Input:} An instance of $PSRC$, period $t$, a partial (in)feasible solution $(L, \mathbf{n}_t, \mathbf{x}_t, \mathbf{z}_t, \mathbf{u}_t)$ \\ 
\hline
\vspace{-4mm}\\
0. Set $\mathbf{u}_t = \mathbf{0}$. \\ 
\hspace{8pt} Let $\mathcal{L}$ be a list of sensors and sinks. Add $S-$many sinks to $\mathcal{L}$ and label them. \\
1. \textbf{While} $\mathcal{L}$ is not empty \\
2. \hspace{10pt} \textbf{If} the first element of $\mathcal{L}$ is a sink (say $(j, 0)$), \textbf{Then} \\
 \hspace{30pt} add unlabeled active sensors $(i, l)$ that communicate \\
 \hspace{30pt} with the sink $(j, 0)$, i.e., $b_{ilj} = 1$, to list $\mathcal{L}$.  Set $u_{jilt} = 1$. \\
 \hspace{30pt} Remove sink $(j, 0)$ from list $\mathcal{L}$. \\
3. \hspace{10pt} \textbf{Else} /* the first element of $\mathcal{L}$ is a sensor (say $(j, k)$) */ \\
 \hspace{30pt} add unlabeled active sensors $(i, l)$ that communicate  \\
 \hspace{30pt} with the active sensor $(j, k)$, i.e., $b_{ilj} = 1$, to list $\mathcal{L}$. Set $u_{vilt} \leftarrow u_{vjkt}$ for all $v \in \mathbf{N}$. \\
 \hspace{30pt} Remove sensor $(j, k)$ from list $\mathcal{L}$.  \\
4. \hspace{10pt} \textbf{End If}\\
5. \textbf{End While} \\
6.  \textbf{If} connectivity is not satisfied,  \textbf{Then} \\
7.  \hspace{10pt} \textbf{While} there is $CoEP_{jk} > 0$ with sufficient $E^{rem}_{jkt}$ for some sensor $(j, k)$ \\
 \hspace{22pt} activate ($z_{jkt} = 1$) the standby sensor $(j, k)$ with the highest $CoEP_{jk}$. \\
8. \hspace{10pt} \textbf{End While} \\
9.  \hspace{10pt} \textbf{While} there is a sensor  $(j, k)$ with the highest $CoCR_{jk} > 0$ \\
10.  \hspace{20pt} \textbf{If} budget is not enough for the sensor $(j, k)$, \textbf{Then} order standby sensors  \\
\hspace{38pt} in $[0, t]$ in nonincreasing cost and remove them until enough budget is obtained. \\
11.  \hspace{20pt} \textbf{If} budget is enough,\\
12. \hspace{20pt}  \textbf{Then} deploy ($x_{jk} = 1$) and activate ($z_{jkt} = 1$) the sensor $(j, k)$ with the highest $CoCR_{jk}$. \\
\hspace{38pt} Pick a sensor $(i, l)$ with $b_{jki} = 1$, and set  $u_{vjkt} \leftarrow u_{vilt}$ for all $v \in \mathbf{N}$. \\
13.  \hspace{20pt} \textbf{Else} $n_t = 0, L = t - 1$ and Stop. \\
14. \hspace{20pt} \textbf{End If}\\
15. \hspace{10pt} \textbf{End While} \\
16. \textbf{End If}\\
\hline
\vspace{-4mm}\\
\textbf{Output:}  A solution satisfying the constraints (\ref{connectivity}) and (\ref{feasibleu1}) - (\ref{feasibleu3})
for period $t$. \\  
\hline
\end{tabular}
\label{alg:FeasSinkAssign}
\end{algorithm}
\end{onehalfspace}
\end{minipage}
\end{center}

We satisfy the connectivity (\ref{connectivity}) and the sink assignment (\ref{feasibleu1}) - (\ref{feasibleu3}) constraints in a period $t$ with Algorithm \ref{alg:FeasSinkAssign}. We carry out a breadth--first--search starting from $S-$many sinks to detect the active sensors that can communicate with one of the sinks (Step 2) or another active sensor that has a sink assignment (Step 3). If each active sensor cannot communicate with at least $\alpha-$many active sensors as constraints (\ref{connectivity}) require, we first consider to activate some of the standby sensors $(j, k)$, since this is cost free. 
Among the standby sensors $(j, k)$ having  $E^{rem}_{jkt} \geq (e_{k}^{s} + \zeta_t (e_{k}^{r} + e_{k}^{c}))$ remaining battery energy, we activate the one which has the highest positive \emph{communication energy product}, i.e., $CoEP_{jk}$ (Step 7). Let $(\mathbf{x}_t, \mathbf{z}_t)$ be the partial initial solution obtained with Algorithm \ref{alg:FeasCoverBudget} in period $t$. Then, we calculate the underconnectivity $UC_{il}$ of the active sensors $(i, l)$ with Equation (\ref{UCil}).

\vspace{-3mm}
\begin{equation}
UC_{il} = \max \left\{\left(\alpha z_{ilt}- \underset{\begin{subarray}{c}
  (j, k) \in \mathbf{N} \Cross \mathbf{K} \\
  (j, k) \neq (i, l)
  \end{subarray}}{\sum} b_{ilj}z_{jkt}\right), 0 \right\}, \;\; (i, l)  \in  \mathbf{N} \Cross \mathbf{K} \label{UCil}
\end{equation}

In order to provide connectivity, we prefer to activate the cheapest standby sensor that has the highest remaining battery energy and can communicate with most of the underconnected sensors. Hence, we define $CoEP_{jk}$ for a standby sensor $(j, k)$ in period $t$ as

\vspace{-3mm}
\begin{equation}
CoEP_{jk} = \frac{\left(\underset{\begin{subarray}{c}
(i, l)  \in \mathbf{N}\Cross \mathbf{K}\\ UC_{il} > 0
  \end{subarray}}{\sum}b_{jki}\right)x_{jk}E^{rem}_{jkt}}{c_{jk}}, \;\;   (j, k)  \in \mathbf{N} \Cross \mathbf{K}. \label{CoEPjk}
\end{equation}

If connectivity is not held in Step 12, we consider to deploy ($x_{jk} = 1$) and activate ($z_{jkt} = 1$) the sensor $(j, k)$ with the highest positive \emph{communication cost ratio}, i.e.,  $CoCR_{jk}$ given as 

\vspace{-3mm}
\begin{equation}
CoCR_{jk} = \frac{\left(\underset{\begin{subarray}{c}
(i, l)  \in \mathbf{N}\Cross \mathbf{K}\\ UC_{il} > 0
  \end{subarray}}{\sum}b_{jki}\right)\left(1 - x_{jk}\right)E_{k}}{c_{jk}}, \;\;  (j, k)  \in \mathbf{N} \Cross \mathbf{K} \label{CoCRjk}
\end{equation}
where $E_{k}$ represents the initial battery energy of a type-$k$ sensor.

We continue with the sensor deployment as we can find a sensor that has positive $CoCR_{jk}$ value.  In the insufficient budget case, we remove the sensors that are in standby mode upto period $t$. If we cannot generate the required budget, we update the network lifetime as $L = t - 1$ and stop the algorithm. 


In the last step of CH (Step 8 of Algorithm \ref{alg:Constructive}), we solve $RP(t)$ for each period $t$ within the lifetime $L$ to find the minimum energy consuming sensor--to--sink data flow paths. In $RP(t)$,  $y_{jkil}$ and $g_{jki}$ are continuous variables representing the data flows in period $t$ from sensor $(j, k)$ to sensor $(i, l)$ and from sensor $(j, k)$ to sink $(i, 0)$, respectively. We have a partial solution $(\bar{L}, \mathbf{\bar{n}}, \mathbf{\bar{x}}, \mathbf{\bar{z}}, \mathbf{\bar{u}})$ obtained by applying Algorithms \ref{alg:FeasCoverBudget} and \ref{alg:FeasSinkAssign}. We initialize the remaining energy of a sensor $(j, k)$ in period $t = 1$ as $E^{rem}_{jk1} = E_k$. Let $\varepsilon_{jk}^t$ be the value of $\varepsilon_{jk}$ variable in $RP(t)$, i.e., energy consumption of the sensor  $(j, k)$ in period $t$. Then, we update the battery energy $E^{rem}_{jk,t+1}$ at the beginning of the period $t+ 1$ as 

\vspace{-3mm}
\begin{equation}
E^{rem}_{jk,t+1} = E^{rem}_{jkt} - \varepsilon_{jk}^t, \;\;  (j, k)  \in \mathbf{N} \Cross \mathbf{K}. \label{EnergyUpdate}
\end{equation}

Observe that we can find a feasible solution of $RP(t)$ for each period $t$, since we activate the sensors that have sufficient energy for transmitting maximum possible flow $\zeta_t$ in Algorithms \ref{alg:FeasCoverBudget} and \ref{alg:FeasSinkAssign}. We rewrite the constraints (\ref{energy}), (\ref{flowbalance}), (\ref{sinkflow}), (\ref{feasibleflow}) -- (\ref{flowM}), (\ref{feasiblesinkflow}), (\ref{flowM2}), (\ref{feasibleflow1}) as the constraints (\ref{energyValue}), (\ref{fbalance}), (\ref{sinkflowbalance}), (\ref{feasflow}) -- (\ref{fM}), (\ref{feassinkflow}), (\ref{fM2}), (\ref{feasflow1}), respectively, with $y_{jkil}$ and $g_{jki}$ variables and given $(\mathbf{\bar{x}}, \mathbf{\bar{z}}, \mathbf{\bar{u}})$ values in $RP(t)$. Constraints (\ref{EnergyBound}) bound the energy consumption with $E^{rem}_{jkt}$ which is calculated with the Equations (\ref{EnergyUpdate}). Observe that $\mathbf{w}$ variables are not required in $RP(t)$, since we can represent the constraints (\ref{linear1}) -- (\ref{linear4}) with the constraints (\ref{yufeasibleflow}) and (\ref{gufeasibleflow}). Furthermore, we do not need constraints (\ref{alphaoutflows}), since $RP(t)$ avoids unnecessary flows by minimizing the energy consumption. $RP(t)$ can be solved in polynomial time, since it is a linear programming formulation \cite{K84}.

\textbf{RP($t$)}:
\vspace{-21pt}
\begin{subequations} \label{RPt}
\footnotesize
\singlespace
\begin{align}
\mbox{min} & \;\; \underset{\begin{subarray}{c}
(j, k)  \in \mathbf{N}\Cross \mathbf{K}
  \end{subarray}}{\sum} \varepsilon_{jk} \label{objective2} \\ 
\mbox{s.t.: } & \nonumber \\ 
 & \varepsilon_{jk} = e_k^s \bar{z}_{jkt} + e_k^r \sum_{(i, l)  \in  \mathbf{N} \Cross \mathbf{K}} y_{iljk} + e_k^c\sum_{(i, l)  \in  \mathbf{N} \Cross \mathbf{K}} y_{jkil}  + e_k^c  \sum_{i \in \mathbf{N}}  g_{jki}, \;\;  (j, k) \in \mathbf{N} \Cross \mathbf{K} \label{energyValue}  \\
 & \varepsilon_{jk} \leq E^{rem}_{jkt}, \;\;  (j, k)  \in \mathbf{N} \Cross \mathbf{K} \label{EnergyBound} \\
 & \sum_{(i, l)  \in  \mathbf{N} \Cross \mathbf{K}} y_{iljk} + h_{jk}\bar{z}_{jkt} = \sum_{(i, l)  \in  \mathbf{N} \Cross \mathbf{K}} y_{jkil} +  \sum_{i \in \mathbf{N}}  g_{jki}, \;\;  (j, k) \in \mathbf{N} \Cross \mathbf{K} \label{fbalance} \\
 &\sum_{(j, k)  \in  \mathbf{N} \Cross \mathbf{K}} g_{jki} = \sum_{(j, k)  \in  \mathbf{N} \Cross \mathbf{K}}  h_{jk}\bar{u}_{ijkt}, \;\;  i \in \mathbf{N} \label{sinkflowbalance} \\
 & y_{iljk} \leq M_1b_{ilj}, \;\;   (i, l),  (j, k) \in \mathbf{N} \Cross \mathbf{K} \label{feasflow} \\
 & \sum_{(i, l)  \in  \mathbf{N} \Cross \mathbf{K}} y_{jkil} \leq M_1 \bar{z}_{jkt}, \;\;  (j, k) \in \mathbf{N} \Cross \mathbf{K} \label{fM1} \\ 
 & \sum_{(i, l)  \in  \mathbf{N} \Cross \mathbf{K}} y_{iljk} \leq M_1 \bar{z}_{jkt}, \;\;  (j, k) \in \mathbf{N} \Cross \mathbf{K} \label{fM} \\
 & y_{iljk} \leq M_1\left(1 - \sum_{v\in \mathbf{N}} \frac{|\bar{u}_{vilt} - \bar{u}_{vjkt}|}{2}\right), \;\;(i, l),  (j, k) \in \mathbf{N} \Cross \mathbf{K}  \label{yufeasibleflow} \\
 & g_{ilj} \leq M_2 b_{ilj}, \;\; (i, l) \in \mathbf{N} \Cross \mathbf{K}, j \in \mathbf{N} \label{feassinkflow} \\
 & \sum_{(i, l) \in \mathbf{N} \Cross \mathbf{K}}  g_{ilj} \leq M_2 \bar{x}_{j0},  \;\;  j  \in \mathbf{N} \label{fM2} \\
  & g_{ilj} \leq M_2\left(1 - \sum_{v\in \mathbf{N}} \frac{|\bar{u}_{vilt} - \mathbf{1}_j(v)\bar{x}_{j0}|}{2}\right), \;\;(i, l) \in \mathbf{N} \Cross \mathbf{K}, j \in \mathbf{N} \label{gufeasibleflow} \\
 & y_{ilil} = 0, \;\;   (i, l) \in \mathbf{N} \Cross \mathbf{K} \label{feasflow1} \\
 & y_{iljk} \geq 0, \;\;  (i, l),  (j, k) \in  \mathbf{N} \Cross \mathbf{K}\label{yVar1} \\ 
 & g_{ilj} \geq 0, \;\;  (i, l) \in  \mathbf{N} \Cross \mathbf{K}, j \in  \mathbf{N} \label{gVar1} \\
 & \varepsilon_{jk} \geq 0, \;\;  (j, k) \in  \mathbf{N} \Cross \mathbf{K} \label{eVar}.
\end{align}
\end{subequations}





CH method provides feasibility with respect to coverage (\ref{coverage}) and budget (\ref{budget}) constraints for all periods in the planning horizon $T$ (Step 1 of Algorithm \ref{alg:Constructive}). Then, we assign a sink to each of the active sensors while satisfying connectivity (\ref{connectivity}) constraints for all periods (Step 4 of Algorithm \ref{alg:Constructive}). Lastly, we find the minimum energy consuming paths from sensors to their assigned sinks for all periods (Step 8 of Algorithm \ref{alg:Constructive}). CH consumes most of the budget to cover the network as more periods as possible in Step 1. This strategy decreases the chance of providing connectivity by deploying new sensors in Step 4, which adversely affects the lifetime $L$. From this observation, we propose DH method (see Section \ref{DisjunctiveHeuristic}), which considers feasibility with respect to coverage (\ref{coverage}), budget (\ref{budget}), connectivity (\ref{connectivity}), sink assignment (\ref{feasibleu1}) - (\ref{feasibleu3}) constraints and determination of data routes for each period independently. 
    
\subsubsection{Disjunctive Heuristic} \label{DisjunctiveHeuristic}

We summarize our DH method in Algorithm \ref{alg:Disjunctive}. We first set the active sensors, which do not have sufficient battery energy $E^{rem}_{jkt}$  in a period $t$, to standby mode (Steps 1 - 4). We treat each period individually, and satisfy constraints in the order of coverage (\ref{coverage}), budget (\ref{budget}), connectivity (\ref{connectivity}), sink assignment (\ref{feasibleu1}) - (\ref{feasibleu3}). We remove standby sensors upto the current period when the budget is not enough to deploy new sensors for coverage and connectivity (Steps 11 and 18). 


\begin{center}
\begin{minipage}{0.85\textwidth}
\begin{onehalfspace}
\footnotesize
\begin{algorithm} \textbf{(Disjunctive Heuristic)}\\
\begin{tabular}{ll}
\hline
\vspace{-4mm}\\
\textbf{Input:} An instance of $PSRC$, a vector $(L, \mathbf{n}, \mathbf{x}, \mathbf{z}, \mathbf{u}, \mathbf{y}, \mathbf{g}, \mathbf{w})$  \\ 
\hline
\vspace{-4mm}\\
0.  Set $t = 1$, $L = T$.\\ 
1. \textbf{For Each} $t \leq L$ \\
2. \textbf{For Each} active sensor $(j, k)$ in period $t$ ($z_{jkt} = 1$), calculate $E^{rem}_{jkt}$ \\
3. \hspace{10pt} \textbf{If} $E^{rem}_{jkt}$ is not sufficient, \textbf{Then} $z_{jkt} = 0$. \\
4. \textbf{End For Each} \\
5. \textbf{For Each} $t \leq L$ \\
6. \textbf{If} every node is covered, \textbf{Then} \\
7. \hspace{10pt} \textbf{If} budget is satisfied,  \textbf{Then} \\
8. \hspace{20pt}  \textbf{If}  Algorithm  \ref{alg:FeasSinkAssign} generates a feasible solution subject to constraints (\ref{connectivity}) \\ 
 \hspace{42pt}  and  (\ref{feasibleu1}) - (\ref{feasibleu3})  in period $t$,  \textbf{Then} \\
 \hspace{42pt} Apply Algorithm \ref{alg:PolishSoln} to deactivate ($z_{jkt} = 0$) the sensors \\
 \hspace{42pt} without violating constraints (\ref{coverage}), (\ref{connectivity}) and  (\ref{feasibleu1}) - (\ref{feasibleu3}). \\
 \hspace{41pt} Solve $RP(t)$ to determine the data flows $\mathbf{y}$ and $\mathbf{g}$. \\ 
 \hspace{43pt} $t \leftarrow t + 1$ \\
9. \hspace{20pt} \textbf{Else}  Stop. \\
10. \hspace{15pt} \textbf{End If}\\
11. \hspace{5pt} \textbf{Else} order standby sensors in $[0, t]$ in nonincreasing cost \\
\hspace{22pt} and remove them until budget is satisfied. \\
12. \hspace{17pt} \textbf{If} budget is violated,  \textbf{Then} $n_t = 0, L = t - 1$ and Stop.  \\ 
13. \hspace{17pt} \textbf{Else} apply Steps 8 - 10. \\
14. \hspace{5pt} \textbf{End If}\\
15. \textbf{Else} \textbf{While} there is $CEP_{jk} > 0$ with sufficient $E^{rem}_{jkt}$ for some sensor $(j, k)$ \\
 \hspace{33pt} activate ($z_{jkt} = 1$) the standby sensor $(j, k)$ with the highest $CEP_{jk}$. \\ 
16. \hspace{18pt} \textbf{End While} \\
17.  \hspace{18pt} \textbf{While} there is a sensor  $(j, k)$ with the highest $CCR_{jk} > 0$ \\ 
18.  \hspace{30pt} \textbf{If} budget is not enough for the sensor $(j, k)$, \textbf{Then} order standby sensors  \\
\hspace{48pt} in $[0, t]$ in nonincreasing cost and remove them until enough budget is obtained. \\
19.  \hspace{30pt} \textbf{If} budget is enough, \\
20. \hspace{30pt}  \textbf{Then} deploy ($x_{jk} = 1$) and activate ($z_{jkt} = 1$) the sensor $(j, k)$ with the highest $CCR_{jk}$. \\ 
21.  \hspace{30pt} \textbf{Else} $n_t = 0, L = t - 1$ and Stop. \\
22. \hspace{30pt} \textbf{End If}\\
23. \hspace{18pt} \textbf{End While} \\
24. \hspace{17pt} Apply Steps 8 - 10. \\
25. \textbf{End If}\\
26. \textbf{End For Each} \\
\hline
\vspace{-4mm}\\
\textbf{Output:}  A feasible solution of $PSRC$ with lifetime $L$. \\
\hline
\end{tabular}
\label{alg:Disjunctive}
\end{algorithm}
\end{onehalfspace}
\end{minipage}
\end{center}

Furthermore, we save energy with Algorithm \ref{alg:PolishSoln} by setting the active sensors, which will not harm coverage and connectivity restrictions, to standby mode (Step 8). We start deactivation from the most expensive active sensor, since we can generate more budget if we consider to remove standby sensors for coverage or connectivity in some period. $RP(t)$ finds the minimum energy consuming paths for the active sensors to their assigned sinks in period $t$ (Step 8). We move to the next period if we satisfy the constraints for the current period. This budget and battery energy utilization strategy improves the lifetime $L$ compared to CH as we report in Section \ref{ComputationalResults}.


\begin{center}
\begin{minipage}{0.85\textwidth}
\begin{onehalfspace}
\footnotesize
\begin{algorithm} \textbf{(Deactivating Unnecessary Sensors)}\\
\begin{tabular}{ll}
\hline
\vspace{-4mm}\\
\textbf{Input:} An instance of $PSRC$, period $t$, a partial feasible solution $(L, \mathbf{n}_t, \mathbf{x}_t, \mathbf{z}_t, \mathbf{u}_t)$  \\ 
\hline
\vspace{-4mm}\\
0. Let $isActive = false$. \\
\hspace{7pt} Let $\mathcal{L}$ be the nonincreasing cost order of the active sensors $(j, k)$ in period $t$ ($z_{jkt} = 1$). \\
1. \textbf{For Each} active sensor $(j, k) \in \mathcal{L}$ \\
2. \hspace{10pt} $isActive = false$. \\
3. \hspace{10pt} \textbf{For Each} node $i$ with $a_{ijk} = 1$ \\
4. \hspace{20pt} \textbf{If} constraints (\ref{coverage}) are violated for node $i$, \textbf{Then} $isActive = true$.   \\
5. \hspace{10pt} \textbf{End For Each} \\
6. \hspace{10pt}  \textbf{If} $isActive = false$, \textbf{Then} \\
7. \hspace{10pt} \textbf{For Each} active sensor $(i, l)$ with $b_{jki} = 1$ \\
8. \hspace{20pt} \textbf{If} constraints (\ref{connectivity}) are violated for sensor $(i, l)$, \textbf{Then} $isActive = true$.   \\
9. \hspace{10pt} \textbf{End For Each} \\
10. \hspace{6pt} \textbf{End If} \\
11. \hspace{6pt}  \textbf{If} $isActive = false$, \textbf{Then} \\
12. \hspace{6pt} Deactivate sensor $(j, k)$ in period $t$ ($z_{jkt} = 0$). \\
13. \hspace{6pt} \textbf{For Each} active sensor $(i, l)$ with $b_{jki} = 1$, set $u_{vilt} = null$  for all $v \in \mathbf{N}$. \\
14. \hspace{6pt} \textbf{For Each} active sensor $(i, l)$ without sink assignment \\
15. \hspace{20pt} Pick one of the active sensors $(i', l')$ with $b_{ili'} = 1$ and set $u_{vilt} \leftarrow u_{vi'l't}$ for all $v \in \mathbf{N}$. \\
16. \hspace{6pt} \textbf{End For Each} \\
17. \hspace{6pt} \textbf{End If} \\
18. \textbf{End For Each} \\
\hline
\vspace{-4mm}\\
\textbf{Output:}  Unnecessary active sensors are in standby mode in period $t$. \\
\hline
\end{tabular}
\label{alg:PolishSoln}
\end{algorithm}
\end{onehalfspace}
\end{minipage}
\end{center}

The computational complexity of Algorithm \ref{alg:PolishSoln} is given as $\mathcal{O}(N^{4}K^{3})$.

\subsection{Sink Search Algorithms for $SPSRC$} \label{SinkSearch}

In this section, we describe our search heuristics, i.e., LS and TS, to satisfy constraint (\ref{sinklocate}) while maximizing the network lifetime $L$. These algorithms estimate the lifetime $L$ with a heuristic for $PSRC$ (see Section \ref{PSRCwithSink}) while searching the locations of $S-$many sinks.

\subsubsection{Local Search Heuristic}  \label{LocalSearch}

In LS heuristic, given in Algorithm \ref{alg:LocalSearch}, we explore $N$ candidate locations to deploy $S$ sinks. We assume that $\mathbf{x}_0 = (x_{10}, x_{20}, ..., x_{N0})$ with $x_{j0} \in \{0, 1\}$ is the binary vector of sink locations. We sort $N$ candidate locations in nondecreasing sink deploy cost $c_{j0}$ order. Initially, we locate $S$ sinks to the cheapest nodes. We implement a $PSRC$ heuristic (CH or DH) to obtain an initial network lifetime $L$ with the remaining budget $(B - \sum_{j \in \mathbf{N}} c_{j0} x_{j0})$ (Step 0). 

 At each iteration $iter$, we randomly change the locations of $s \leq S$ sinks. We pick a location $j$ with probability $p(j)$, which is inversely proportional to the sink cost $c_{j0}$ as given in Equation (\ref{probDistribution}). We can have more remaining budget for the $PSRC$ problem by economically locating the sinks with this strategy.

\vspace{-5mm}
\begin{equation}
p(j) = \frac{\underset{\begin{subarray}{c}
 i \neq j \\ i \in \mathbf{N} 
  \end{subarray}}{\sum} c_{i0}}{(N- 1) \underset{\begin{subarray}{c}
 i \in \mathbf{N} 
  \end{subarray}}{\sum}c_{i0}}, \;\; j \in \mathbf{N}. \label{probDistribution} 
\end{equation}

Given the locations $\mathbf{x}_0 $ of $S-$sinks, there are $NS_{s} = \binom{N - S}{s} \binom{S}{s}$ different alternatives to relocate $s-$sinks. 
In LS, we scan $P_s$ percentage of the neighborhood $NS_{s}$ to determine an improving solution (Step 3). Once we relocate the sinks, we compute the network lifetime $\bar{L}$ with the remaining budget through a $PSRC$ heuristic (Step 4). We continue with the search until we complete $iterLim-$many iterations or we cannot update the current lifetime $L$ for $nImpr-$many consecutive iterations.
 
\begin{center}
\begin{minipage}{0.75\textwidth}
\begin{onehalfspace}
\footnotesize
\begin{algorithm} \textbf{(Local Search)}\\
\begin{tabular}{ll}
\hline
\vspace{-4mm}\\
\textbf{Input:} \# of candidate locations $N$, \# of sinks $S$, $iterLim$, $nImpr$  \\ 
\hline
\vspace{-4mm}\\
0. Set $iter = 0$ and $noImpr = 0$. Let $\mathbf{x}_0$ be the vector of sink locations.  \\
\hspace{7pt} Locate $S$ sinks to the cheapest nodes. \\
\hspace{7pt} Let $L$ be the lifetime found by a $PSRC$ heuristic (CH or DH). \\
1. \textbf{While} $iter < iterLim$ and $noImpr < nImpr$ \\
2. \hspace{10pt} \textbf{For Each} $s \leq S$ \\
3. \hspace{20pt} \textbf{For} $NS_sP_s-$many trials    \\
4. \hspace{30pt} Randomly change locations of $s-$sinks. \\
\hspace{42pt} Let $\bar{L}$ be the lifetime found by a $PSRC$ heuristic. \\
5. \hspace{30pt} \textbf{If} $\bar{L} > L$, \textbf{Then} $L \leftarrow \bar{L}$, update $\mathbf{x}_0$. \\
6. \hspace{20pt} \textbf{End For} \\
7. \hspace{10pt} \textbf{End For Each} \\
8.  \hspace{10pt}  \textbf{If} $L$ is improved, \textbf{Then} $noImpr = 0$, \textbf{Else} $noImpr \leftarrow noImpr + 1$. \\
9.  \hspace{10pt} $iter \leftarrow iter + 1$ \\
10. \textbf{End While} \\
\hline
\vspace{-4mm}\\
\textbf{Output:} A feasible solution of $SPSRC$ with lifetime $L$. \\
\hline
\end{tabular}
\label{alg:LocalSearch}
\end{algorithm}
\end{onehalfspace}
\end{minipage}
\end{center}


\subsubsection{Tabu Search Heuristic} \label{TabuSearch}

TS, given in Algorithm \ref{alg:TabuSearch}, aims to visit as distict parts of the solution space as possible by forbidding to revisit the recent $tabuTenure-$many solutions from the solution space of sink locations \cite{GP05}. Similar to LS, we locate $S$ sinks to the cheapest nodes initially. We swap $s-$sinks ($s \leq S$) randomly using the probability density function in Equation (\ref{probDistribution}) to move another solution (Steps 2 - 6). 

We implement a $PSRC$ heuristic (CH or DH) to determine the network lifetime $\bar{L}$ with the current sink locations (Step 4). In Step 5, we add improving solutions to $tabuList$, which stores recent $tabuTenure-$many solutions. That is as we add a new solution to $tabuList$, we remove the oldest solution from $tabuList$. The algorithm stops after $iterLim-$many iterations or $nImpr-$many consecutive nonimproving iterations.


\begin{center}
\begin{minipage}{0.75\textwidth}
\begin{onehalfspace}
\footnotesize
\begin{algorithm} \textbf{(Tabu Search)}\\
\begin{tabular}{ll}
\hline
\vspace{-4mm}\\
\textbf{Input:} \# of candidate locations $N$, \# of sinks $S$, $iterLim$, $nImpr$, $tabuTenure$  \\ 
\hline
\vspace{-4mm}\\
0. Set $t = 0$, $noImpr = 0$ and $tabuList = \emptyset$. Let $\mathbf{x}_0$ be the vector of sink locations.  \\
\hspace{7pt} Locate $S$ sinks to the cheapest nodes. Add $\mathbf{x}_0$ to $tabuList$.\\
\hspace{7pt} Let $L$ be the lifetime found by a $PSRC$ heuristic (CH or DH). \\
1. \textbf{While} $t < iterLim$ and $noImpr < nImpr$ \\
2. \hspace{10pt} \textbf{For Each} $s \leq S$ \\
3. \hspace{20pt} \textbf{For} $NS_sP_s-$many trials    \\
4. \hspace{30pt} Randomly change locations of $s-$sinks to obtain a nontabu vector. \\
\hspace{42pt} Let $\bar{L}$ be the lifetime found by a $PSRC$ heuristic. \\
5. \hspace{30pt} \textbf{If} $\bar{L} > L$, \textbf{Then} $L \leftarrow \bar{L}$, update $\mathbf{x}_0$ and add to $tabuList$. \\
6. \hspace{20pt} \textbf{End For} \\
7. \hspace{10pt} \textbf{End For Each} \\
8.  \hspace{10pt}  \textbf{If} $L$ is improved, \textbf{Then} $noImpr = 0$, \textbf{Else} $noImpr \leftarrow noImpr + 1$. \\
9.  \hspace{10pt} $t \leftarrow t + 1$ \\
10. \textbf{End While} \\
\hline
\vspace{-4mm}\\
\textbf{Output:} A feasible solution of $SPSRC$ with lifetime $L$. \\
\hline
\end{tabular}
\label{alg:TabuSearch}
\end{algorithm}
\end{onehalfspace}
\end{minipage}
\end{center}

\section{Computational Results} \label{ComputationalResults}


The computations have been carried out on a computer with 2.0 GHz Intel Xeon E5-2620 processor and 46 GB of RAM working under Windows Server 2012 R2 operating system. In computational experiments, we used CPLEX 12.7.0 to solve $RP(t)$ model for period $t$ in CH and DH algorithms. We implement all algorithms in the C++ programming language. 

We summarize the computational parameters in Table \ref{tab:locp}.  In our experiments, we consider $n \times n$ square grid region to be monitored. That is, we try eight different network sizes from $N = 16$ to $N = 225$. Each node $i$ in the region has a coverage requirement $f_i = 2$. We maximize the network lifetime $L$ within a planning horizon $T = 400$. We provide connectivity if each active sensor communicates with at least $\alpha = 1$ active sensor in a period. We conduct experiments with three budget $B$ levels, i.e., low, medium and high, which are calculated as in Equations (\ref{budgetLevels}). As an example with the high budget, we can deploy type--1 sensors to 25\% of the nodes, type--2 sensors to 75\% of the nodes and we can locate $S$ sinks of average cost. 

\begin{subequations}  \label{budgetLevels}
\begin{align}
& B_{low} =  \frac{S}{N} \sum_{j \in \mathbf{N}} c_{j0} + 0.75 \sum_{j \in \mathbf{N}} c_{j1} + 0.25  \sum_{j \in \mathbf{N}} c_{j2}, \\
& B_{medium} = \frac{S}{N} \sum_{j \in \mathbf{N}} c_{j0} + 0.50 \sum_{j \in \mathbf{N}} c_{j1} + 0.50 \sum_{j \in \mathbf{N}} c_{j2},    \\
& B_{high} =  \frac{S}{N} \sum_{j \in \mathbf{N}} c_{j0} + 0.25 \sum_{j \in \mathbf{N}} c_{j1} + 0.75  \sum_{j \in \mathbf{N}} c_{j2},  
\end{align}
\end{subequations}

We assume a sink is of type--0 and we have $K = 2$ different sensor types for sensing the region and transmitting the data packets. There is a cost $c_{jk}$ to deploy a sensor or a sink at node $j$, which we randomly determine within the ranges given in Table \ref{tab:locp}. We take a period length as 12 hours and we assume that every half an hour a data packet is generated by a sensor. Hence, a type$-k$ sensor can generate $h_{jk} = 24$ data packets in a period. Sensing $r_k^s$ and communicating $r_k^c$ ranges of a type--2 sensor are twice as large as the ones of a type--1 sensor. A sink (type--0) cannot sense or communicate, and we assume it does not consume energy. We determine the values of the sensor parameters $e_k^s, e_k^r, e_k^c$ and $E_k$ based on the experimental results for a Mica2 mote studied by Calle and Kabara \cite{CK06}. We experiment for three initial battery energy $E_k$ levels, i.e., low, medium and high. At high energy level, the battery of a type$-k$ sensor is full. The medium energy level is 2/3 of the full battery energy and 1/3 of the full battery energy refers to the low energy level.





\vspace{-3mm}
\begin{table}[!h]
\captionof{table}{List of computational parameters}
    \label{tab:locp}
\line(1,0){430}
			\vspace{-0.2in}
\begin{onehalfspace}
\footnotesize
\begin{multicols}{2}
\hspace{5mm}
\begin{tabular}{l l p{1.0in}}
\multicolumn{3}{l}{\textit{Parameters}} \\
&  \\
$N$ & \multicolumn{2}{l}{16, 25, 36, 49, 64,}\\
& \multicolumn{2}{l}{ 81, 100, 225} \\
$K$ & 2 \\
$T$ & 400 \\
$\alpha$ & 1 \\
$f_i$ & \multicolumn{2}{l}{2 for all $i \in \mathbf{N}$} \\
$B$ &\multicolumn{2}{l}{ low, medium, high} \\
&  \\
\multicolumn{3}{l}{\textit{Sink Search Parameters}} \\
& &  \\
& $S$ & 2, 3 \\
& $iterLim$ & 100 \\ 
& $nImpr$ & 20 \\ 
& $tabuTenure$ & 10 \\ 
& & $s = 1$ \ $s = 2$ \ $s = 3$ \\
\cline{3-3}
$P_s (\%)$ &  LS & \hspace{1mm} 20 \hspace{4mm}  40 \hspace{4mm} 40 \\ 
 & TS & \hspace{1mm} 100 \hspace{3mm} 20 \hspace{4mm} 10 \\
Time Limit & 3600 secs & \\
\end{tabular}

\hspace{-5mm}
			\begin{tabular}{l p{0.3in} c c c}
				\multicolumn{3}{l}{\textit{Sensor Specifications}}  \\
				 & \\
 & &  $k = 0$ & $k = 1$ & $k = 2$  \\
\cline{3-5}
& $c_{jk}$ & (10, 15) &  (1, 10) & ($c_{j1}$, $c_{j1}$ + 5) \\ 
& $h_{jk}$  & 0 & 24 & 24 \\ 
& $r_k^s$ & 0 & 1 & 2 \\
& $r_k^c$ & 0 & 1.5 & 3 \\
& $e_k^s$  & 0 & 744 & 744 \\ 
& $e_k^r$  & 0 & 0.01 & 0.01 \\
& $e_k^c$  & 0 & 0.013 & 0.018 \\
 &  low  & $\infty$ & 19200 & 28800 \\
$E_k$ & medium & $\infty$ & 38400 & 57600 \\
  & high & $\infty$ & 57600  & 86400  \\
			\end{tabular}			
	\end{multicols}	
\end{onehalfspace}
\vspace{-0.2in}
\line(1,0){430}
\end{table}

\newpage
\subsection{Computational Results for $PSRC$}

In this section, we give the computational results for our feasible solution generation algorithms CH (see Section \ref{ConstructiveHeuristic}) and DH (see Section \ref{DisjunctiveHeuristic}). We randomly locate $S$ sinks in the network and solve for the lifetime $L$ of the corresponding $PSRC$ problem. 

We investigate the sensitivity of our CH and DH methods to the number of sinks ($S= 2$ or 3), budget level (low, medium, high), energy level  (low, medium, high), and network size (from $N = 16$ to 225). For a given set of parameters $(S, \text{Budget}, \text{Energy}, N)$, we randomly generate 10 instances  and report the average values. In the following tables, the column ``$L$" is the network lifetime and ``CPU (secs)" is the computational time in seconds.  

\begin{table}[!h]
\centering
\tiny
\caption{Computational results for CH}
\begin{tabular}{ccccccccccccccc} 
\hline
& & \multicolumn{6}{c}{$S = 2$} & & \multicolumn{6}{c}{$S = 3$} \\ \cline{3 - 8} \cline{10 - 15} \vspace{1mm}
& Budget & \multicolumn{2}{c}{Low} & \multicolumn{2}{c}{Medium} & \multicolumn{2}{c}{High} & &  \multicolumn{2}{c}{Low} & \multicolumn{2}{c}{Medium} & \multicolumn{2}{c}{High} \\ \cline{3 - 8} \cline{10 - 15}
\multirow{2}{*}{Energy} & \multirow{2}{*}{$N$} & \multirow{2}{*}{$L$} & CPU & \multirow{2}{*}{$L$}  & CPU &  \multirow{2}{*}{$L$} & CPU & &  \multirow{2}{*}{$L$} & CPU &  \multirow{2}{*}{$L$}  & CPU &  \multirow{2}{*}{$L$}  & CPU \\
& & & (secs) & & (secs)  & &  (secs) & & & (secs) & &  (secs)  & &  (secs) \\ 
 \hline
\multirow{8}{*}{Low} & 16 & 79.8 & 1.05 & 79.8 & 1.01 & 79.8 & 1.06 & & 79.8 & 0.93 & 79.8 & 0.94 & 79.8 & 0.94 \\
& 25 & 83.6 & 2.28 & 83.6 & 2.29 & 83.6 & 2.26  & & 83.6 & 2.26 & 83.6 & 2.32 & 83.6 & 2.33 \\
& 36 & 64.8 & 3.87 & 64.8 & 3.81 & 64.8 & 3.88 & & 64.8 & 3.82 & 64.8 & 3.85 & 64.8 & 3.85 \\
& 49 & 74.0 & 8.99 & 74.0 & 9.04 & 74.0 & 8.82 & & 74.0 & 8.84 & 74.0 & 9.03 & 74.0 & 9.02 \\
& 64 & 72.0 & 15.77 & 72.0 & 15.69 & 72.0 & 15.78 & & 72.0 & 16.02 & 72.0 & 15.81 & 72.0 & 16.29 \\
& 81 & 75.6 & 29.63 & 75.6 & 29.42 & 75.6 & 29.55 & & 75.6 & 29.75 & 75.6 & 29.82 & 75.6 & 29.63 \\
& 100 & 75.6 & 52.11 & 75.6 & 52.30 & 75.6 & 52.01 & & 75.6 & 52.61 & 75.6 &52.14 & 75.6 & 52.32 \\
& 225 & 62.3 & 476.95 & 62.8 & 481.37 & 62.8 & 500.41 & & 62.8 & 479.64 & 62.8 & 479.87 & 62.8 & 482.89 \\ \hline
\multirow{8}{*}{Medium} & 16 & 159.6 & 1.91 & 159.6 & 1.89 & 159.6 & 1.89 &  & 159.6 & 1.87 & 159.6 & 1.87 & 159.6 & 1.89 \\
& 25 & 167.2 & 4.58 & 167.2 & 4.52 & 167.2 & 4.59 & &  167.2 & 4.54 & 167.2 & 4.56 & 167.2 & 4.51 \\
& 36 & 133.1 & 7.91 & 133.1 & 7.87 & 133.1 & 7.95 & &  133.1 & 8.15 & 133.1 & 7.97 & 133.1 & 7.92 \\
& 49 & 152.0 & 18.20 & 152.0 & 18.03 & 152.0 & 18.18 &  & 152.0 & 17.95 & 152.0 & 17.89 & 152.0 & 17.89 \\
& 64 & 150.0 & 33.24 & 150.0 & 32.65 & 150.0 & 33.26 & &  150.0 & 33.16 & 150.0 & 32.99 & 150.0 & 33.44 \\
& 81 & 157.2 & 61.24 & 157.2 & 61.66 & 157.2 & 61.52 & &  157.2 & 61.40 & 157.2 & 61.49 & 157.2 & 62.08 \\
& 100 & 155.4 & 107.60 & 155.4 & 107.51 & 155.4 & 107.28 & &  155.4 & 108.25 & 155.4 & 107.79 & 155.4 & 107.48 \\
& 225 & 138.2 & 1156.94 & 139.3 & 1119.86 & 139.3 & 1114.54 &  & 139.3 & 1068.99 & 139.3 & 1074.83 & 139.3 & 1071.55 \\ \hline
\multirow{8}{*}{High} & 16 & 241.5 & 2.78 & 241.5 & 2.80 & 241.5 & 2.86 & & 241.5 & 2.85 & 241.5 & 2.88 & 241.5 & 2.82 \\
& 25 & 253.0 & 6.90 & 253.0  & 6.91 & 253.0  & 6.84 & & 253.0  & 6.91 & 253.0 & 7.05 & 253.0 & 6.89 \\
& 36 & 201.4 & 11.89 & 201.4 & 11.87 & 201.4 & 11.88 & & 201.4 & 11.94 & 201.4 & 11.88 & 201.4 & 11.85 \\
& 49 & 228.0 & 26.92 & 228.0 & 27.00 & 228.0 & 27.07 & & 228.0 & 26.89 & 228.0 & 27.00 & 228.0 & 26.89 \\
& 64 & 228.0 & 50.24 & 228.0 & 49.56 & 228.0 & 50.65 & & 228.0 & 49.33 & 228.0 & 49.97 & 228.0 & 50.73 \\
& 81 & 236.8 & 92.57 & 236.8 & 93.25 & 236.8 & 94.48 & & 236.8 & 92.66 & 236.8 & 92.57 & 236.8 & 93.17 \\
& 100 & 237.3 & 163.71 & 237.3 & 163.50 & 237.3 & 162.72 & & 237.3 & 167.31 & 237.3 & 165.55 & 237.3 & 163.28 \\
& 225 & 213.8 & 1638.72 & 215.4 & 1648.74 & 215.4 & 1651.97 & & 215.4 & 1651.01 & 215.4 & 1654.32 & 215.4 & 1648.95 \\
\hline
\end{tabular}
\label{tab:CH_1Iteration}
\end{table}

As we summarize in Table \ref{tab:CH_1Iteration}, lifetime $L$ increases as the energy level gets higher with CH method. Lifetime $L$ improves when we increase the budget level from low to medium for $S = 2$, $N = 225$ both in medium (from 138.2 to 139.3) and high (from 213.8 to 215.4) energy levels. Besides,  at low budget level and $N = 225$, we have larger lifetime $L$ when we have $S = 3$ instead of $S = 2$ for both medium (from 138.2 to 139.3) and high (from 213.8 to 215.4)  energy levels. It seems that, for a given budget and energy level, lifetime $L$ gets smaller as the network size $N$ increases, since satisfying coverage and connectivity restrictions requires more budget and energy resources. 

\begin{table}[!h]
\centering
\tiny
\caption{Computational results for DH}
\begin{tabular}{ccccccccccccccc} 
\hline
& & \multicolumn{6}{c}{$S = 2$} & & \multicolumn{6}{c}{$S = 3$} \\ \cline{3 - 8} \cline{10 - 15} \vspace{1mm}
& Budget & \multicolumn{2}{c}{Low} & \multicolumn{2}{c}{Medium} & \multicolumn{2}{c}{High} & &  \multicolumn{2}{c}{Low} & \multicolumn{2}{c}{Medium} & \multicolumn{2}{c}{High} \\ \cline{3 - 8} \cline{10 - 15}
\multirow{2}{*}{Energy} & \multirow{2}{*}{$N$} & \multirow{2}{*}{$L$} & CPU & \multirow{2}{*}{$L$}  & CPU &  \multirow{2}{*}{$L$} & CPU & &  \multirow{2}{*}{$L$} & CPU &  \multirow{2}{*}{$L$}  & CPU &  \multirow{2}{*}{$L$}  & CPU \\
& & & (secs) & & (secs)  & &  (secs) & & & (secs) & &  (secs)  & &  (secs) \\ 
 \hline
\multirow{8}{*}{Low} & 16 & 83.6 & 1.04 & 83.6 & 1.08 & 83.6 & 1.08 & & 83.6 & 1.01 & 83.6 & 1.05 & 83.6 & 1.19 \\
& 25 & 87.4 & 2.55 & 87.4 & 2.49 & 87.4 & 2.54 & & 87.4 & 2.51 & 87.4 & 2.45 & 87.4 & 2.58 \\
& 36 & 85.1 & 5.35 & 85.1 & 5.25 & 85.1 & 5.58 & & 85.1 & 5.01 & 85.1 & 5.44 & 85.1 & 4.94 \\
& 49 & 88.8 & 10.71 & 92.5 & 10.76 & 96.2 & 11.31 & & 88.8 & 10.30 & 92.5 & 10.85 & 96.2 & 11.34 \\
& 64 & 79.2 & 17.37 & 79.2 & 17.19 & 79.2 & 17.19 & & 79.2 & 17.26 & 79.2 & 17.23 & 79.2 & 16.96 \\
& 81 & 82.8 & 31.84 & 82.8 & 32.18 & 82.8 & 31.95 & & 82.8 & 31.90 & 82.8 & 32.08 & 82.8 & 31.70 \\
& 100 & 90.0 & 61.29 & 90.0 & 61.23 & 90.0 & 60.95 & & 90.0 & 61.42 & 90.0 & 61.70 & 90.0 & 61.52 \\
& 225 & 73.6 & 564.39 & 73.6 & 559.04 & 73.6 & 562.04 & & 73.6 & 568.90 & 73.6 & 570.27 & 73.6 & 567.52 \\ \hline
\multirow{8}{*}{Medium} & 16 & 167.2 & 2.17 & 167.2 & 2.00 & 167.2 & 2.05 & & 167.2 & 2.27 & 167.2 & 2.16 & 167.2 & 2.27 \\
& 25 & 174.8 & 5.16 & 174.8 & 5.09 & 174.8 & 4.99 & & 174.8 & 4.72 & 174.8 & 4.68 & 174.8 & 4.81 \\
& 36 & 174.8 & 10.90 & 174.8 & 10.79 & 174.8 & 10.92&  & 174.8 & 10.19 & 174.8 & 10.48 & 174.8 & 10.11 \\
& 49 & 182.4 & 21.61 & 190.0 & 22.16 & 197.6 & 23.26 & & 182.4 & 21.14 & 190.0 & 22.15 & 197.6 & 23.35 \\
& 64 & 165.0 & 35.54 & 165.0 & 36.10 & 165.0 & 35.50 & &  165.0 & 35.67 & 165.0 & 36.02 & 165.0 & 35.20 \\
& 81 & 172.5 & 67.09 & 172.5 & 67.40 & 172.5 & 66.99 & & 172.5 & 66.78 & 172.5 & 66.45 & 172.5 & 66.46 \\
& 100 & 185.0 & 126.25 & 185.0 & 125.22 & 185.0 & 126.13 & & 185.0 & 126.94 & 185.0 & 127.49 & 185.0 & 126.67 \\
& 225 & 163.3 & 1310.15 & 163.3 & 1305.04 & 163.3 & 1257.06 & & 163.3 & 1256.47 & 163.3 & 1258.67 & 163.3 & 1261.20 \\ \hline
\multirow{8}{*}{High} & 16 & 253.0 & 2.94 & 253.0 & 2.85 & 253.0 & 3.08 & & 253.0 & 3.25 & 253.0 & 3.13 & 253.0 & 3.00 \\
& 25 & 264.5 & 7.90 & 264.5 & 7.80 & 264.5 & 7.73 & & 264.5 & 7.23 & 264.5 & 7.65 & 264.5 & 7.61 \\
& 36 & 264.5 & 16.21 & 264.5 & 15.66 & 264.5 & 15.61 & & 264.5 & 15.55 & 264.5 & 15.32 & 264.5 & 15.46 \\
& 49 & 273.6 & 31.60 & 285.0 & 33.24 & 296.4 & 35.61 & & 273.6 & 31.77 & 285.0 & 33.13 & 296.4 & 34.69 \\
& 64 & 250.8 & 54.70 & 250.8 & 54.36 & 250.8 & 54.06 & & 250.8 & 53.90 & 250.8 & 53.96 & 250.8 & 54.17 \\
& 81 & 259.9 & 101.71 & 259.9 & 100.85 & 259.9 & 100.24 & & 259.9 & 100.08 & 259.9 & 99.40 & 259.9 & 100.70 \\
& 100 & 282.5 & 192.46 & 282.5 & 191.72 & 282.5 & 191.96 & & 282.5 & 192.71 & 282.5 & 197.72 & 282.5 & 193.90 \\
& 225 & 253.0 & 1952.69 & 253.0 & 1955.53 & 253.0 & 2004.21 & & 253.0 & 2110.47 & 253.0 & 1951.13 & 253.0 & 1952.91 \\
\hline
\end{tabular}
\label{tab:DH_1Iteration}
\end{table}

We give the performance of DH method in Table \ref{tab:DH_1Iteration}. We observe that, DH gives better $L$ values than CH for all set of parameters $(S, \text{Budget}, \text{Energy}, N)$. This difference occurs since DH uses energy and budget resources economically by considering coverage and connectivity  restrictions of each period independently. On the other hand, CH consumes most of the budget and energy to satisfy coverage constraints as more periods as possible. Hence, we may not have sufficient remaining budget to deploy new sensors if the connectivity restrictions are not satisfied in a period.  

The results in Table \ref{tab:DH_1Iteration} show that lifetime $L$ is not improved as we deploy more sinks in the network. We expect to see the effect of the number of sinks $S$ on the lifetime $L$ for larger networks. In large networks energy consumption in routing becomes dominant, since there will be more sensors on the path from a sensor to its assigned sink. One can observe for $N = 49$ that increasing budget from low to medium and then to high improves the lifetime $L$ for all energy levels. For example at medium energy level, we have $L = 182.4$ at low budget level, which becomes 190 for medium budget level and  197.6 for high budget level. 


We also experiment for the performance of CPLEX 12.7.0  within 3600 seconds time limit.  We implement $SPSRC$ formulation with known sink locations ($SPSRC$ formulation reduces to $PSRC$). As we report in Table \ref{tab:CPLEX}, the best known solution that CPLEX 12.7.0 can find has lifetime $L = 1$ within the time limit even for the instances of $(S = 2, \text{High Budget}, \text{High Energy}, N = 16)$ when $T = 400$. Besides, CPLEX cannot improve the trivial upper bound UB = 400. 

\begin{table}[!h]
\centering
\footnotesize
\caption{Computational results for CPLEX for $N = 16$}
\begin{tabular}{ccccccccc} 
\hline
& & \multicolumn{3}{c}{$S = 2$} & & \multicolumn{3}{c}{$S = 3$} \\ \cline{3 - 5} \cline{7 - 9} \vspace{1mm}
\multirow{2}{*}{Energy} & \multirow{2}{*}{Budget} & \multirow{2}{*}{$L$} & \multirow{2}{*}{UB} & CPU & &  \multirow{2}{*}{$L$} & \multirow{2}{*}{UB}  & CPU  \\
& & & & (secs) & & & & (secs)   \\ 
 \hline
\multirow{3}{*}{Low} & Low & 1 & 400 & $time$ & & 1 & 400 & $time$ \\
& Medium & 1 & 400 & $time$ & & 1 & 400 & $time$\\
& High & 1 & 400 & $time$ & & 1 & 400 & $time$  \\ \hline
\multirow{3}{*}{Medium} & Low & 1 & 400 & $time$ & & 1 & 400 & $time$  \\
& Medium & 1 & 400 & $time$ & & 1 & 400 & $time$ \\
& High & 1 & 400 & $time$ & & 1 & 400 & $time$ \\ \hline
\multirow{3}{*}{High} & Low & 1 & 400 & $time$ & & 1 & 400 & $time$ \\
& Medium & 1 & 400 & $time$ & & 1 & 400 & $time$ \\
& High & 1 & 400 & $time$ & & 1 & 400 & $time$ \\ 
\hline
\end{tabular}
\label{tab:CPLEX}
\end{table}

\subsection{Computational Results for $SPSRC$}

We summarize the computational results for our sink location search algorithms LS (see Section \ref{LocalSearch}) and TS (see Section \ref{TabuSearch}) in this section. As we note at Step 0 of Algorithms \ref{alg:LocalSearch} and \ref{alg:TabuSearch}, LS and TS require a heuristic for $PSRC$. We utilize our DH method in LS and TS algorithms to generate a feasible solution of $PSRC$, since DH provides higher $L$ values than CH. 

We generate 10 random instances for each parameter set $(S, \text{Budget}, \text{Energy}, N)$ and report the average results. We impose 3600 seconds of time limit to our search algorithms. We bound the number of iterations in LS and TS with $iterLim = 100$. Besides, we stop the search if we cannot update the lifetime $L$ for $nImpr = 20$ consecutive iterations. The length of the tabu list, i.e., $tabuTenure$, in TS is 10. As given in Table \ref{tab:locp}, we scan $P_s$ percentage of the $s-$swap neighborhood in LS and TS. For example, we scan $P_1 = 20\%$ of the 1--swap neighborhood in LS, whereas  $P_1 = 100\%$ in TS.

We give the results for LS method in Table \ref{tab:LS_DH}. Comparison of Tables \ref{tab:DH_1Iteration} and \ref{tab:LS_DH} shows that the lifetime $L$ improves when we carry out search for the locations of the sinks in the network. For example, LS prolongs the average lifetime $L$ by 0.6 compared to  DH for the instance $(S = 2, N = 100)$ with medium energy and low budget ($L = 185.0$ in Table \ref{tab:DH_1Iteration}  and $L = 185.6$ in Table \ref{tab:LS_DH}).

\begin{table}[h]
\centering
\tiny
\caption{Computational results for LS with DH}
\begin{tabular}{ccccccccccccccc} 
\hline
& & \multicolumn{6}{c}{$S = 2$} & & \multicolumn{6}{c}{$S = 3$} \\ \cline{3 - 8} \cline{10 - 15} \vspace{1mm}
& Budget & \multicolumn{2}{c}{Low} & \multicolumn{2}{c}{Medium} & \multicolumn{2}{c}{High} & &  \multicolumn{2}{c}{Low} & \multicolumn{2}{c}{Medium} & \multicolumn{2}{c}{High} \\ \cline{3 - 8} \cline{10 - 15}
\multirow{2}{*}{Energy} & \multirow{2}{*}{$N$} & \multirow{2}{*}{$L$} & CPU & \multirow{2}{*}{$L$}  & CPU &  \multirow{2}{*}{$L$} & CPU & &  \multirow{2}{*}{$L$} & CPU &  \multirow{2}{*}{$L$}  & CPU &  \multirow{2}{*}{$L$}  & CPU \\
& & & (secs) & & (secs)  & &  (secs) & & & (secs) & &  (secs)  & &  (secs) \\ 
 \hline
\multirow{8}{*}{Low}  & 16 & 84.0 & 4.15 & 84.1 & 4.92 & 84.1 &  4.22 & & 84.0 & 14.59 & 84.1 & 14.57 & 84.2 & 14.61 \\
 & 25 & 87.5 & 23.42 & 87.6 & 23.37 & 87.8 & 24.71 & & 87.8 & 169.47 & 87.6 & 169.36 & 88.1 & 169.27 \\
 & 36 & 85.4 & 128.22 & 85.7 & 128.29 & 85.7 & 128.69 & & 85.3 & 1443.70 & 85.4 & 1455.19 & 85.7 & 1458.50 \\
 & 49 & 88.8 & 1239.15 & 92.6 & 1245.04 & 96.8 & 1241.26 & & 92.5 & $time$ & 96.4 & $time$ & 96.9 & $time$ \\
 & 64 & 79.2 & 2399.15 & 79.6 & 2403.04 & 80.7 & 2401.26 & & 79.3 & $time$ & 80.1 & $time$ & 81.0 & $time$ \\
 & 81 & 83.0 & $time$ & 83.9 & $time$ & 84.7 & $time$ & & 83.0 & $time$ & 83.7 & $time$ & 84.6 & $time$ \\
 & 100 & 90.8 & $time$ & 91.2 & $time$ & 91.5 & $time$  & & 90.6 & $time$ & 91.0 & $time$ & 91.4 & $time$ \\
 & 225 & 75.6 & $time$ & 75.7 & $time$ & 76.0 & $time$ & &  75.6 & $time$ & 75.7 & $time$ & 75.9 & $time$ \\  \hline
\multirow{8}{*}{Medium}  & 16 & 167.6 & $time$ & 167.7 & $time$ & 167.7 & $time$  & & 167.7 & $time$ & 167.7 & $time$ & 167.8 & $time$ \\
 & 25 & 174.9 & $time$ & 175.0 & $time$ & 175.2 & $time$ & & 175.2 & $time$ & 175.3 & $time$ & 175.5 & $time$ \\
 & 36 & 175.1 & $time$ & 175.1 & $time$ & 175.1 & $time$ & & 174.9 & $time$ & 175.0 & $time$ & 175.1 & $time$ \\
 & 49 & 182.4 & $time$ & 190.1 & $time$ & 198.1 & $time$ & & 190.0 & $time$ & 197.8 & $time$ & 198.0 & $time$ \\
 & 64 & 165.0 & $time$ & 165.4 & $time$ & 166.3 & $time$ & & 165.1 & $time$ & 165.9 & $time$ & 166.8 & $time$ \\
 & 81 & 172.7 & $time$ & 173.5 & $time$ & 174.1 & $time$ & & 172.7 & $time$ & 173.3 & $time$ & 174.0 & $time$ \\
 & 100 & 185.6 & $time$ & 186.2 & $time$ & 186.4 & $time$ & & 185.6 & $time$ & 186.0 & $time$ & 186.4 & $time$ \\
 & 225 & 165.3 & $time$ & 165.4 & $time$ & 165.7 & $time$ & &  165.3 & $time$ & 165.4 & $time$ & 165.6 & $time$ \\  \hline
\multirow{8}{*}{High}  & 16 & 253.4 & $time$ & 253.5 & $time$ & 253.5 & $time$ & & 253.4 & $time$ & 253.4 & $time$ & 253.5 & $time$ \\
 & 25 & 264.6 & $time$ & 264.7 & $time$ & 264.9 & $time$ & & 264.8 & $time$ & 265.0 & $time$ & 265.2 & $time$ \\
 & 36 & 264.8 & $time$ & 264.8 & $time$ & 264.8 & $time$ & & 264.6 & $time$ & 264.7 & $time$ & 264.8 & $time$ \\
 & 49 & 273.6 & $time$ & 285.1 & $time$ & 296.8 & $time$ &  & 285.0 & $time$ & 296.6 & $time$ & 296.7 & $time$ \\
 & 64 & 250.8 & $time$ & 251.2 & $time$ & 251.5 & $time$ & & 250.9 & $time$ & 251.3 & $time$ & 251.8 & $time$ \\
 & 81 & 260.1 & $time$ & 260.9 & $time$ & 261.5 & $time$ & & 260.1 & $time$ & 260.7 & $time$ & 261.4 & $time$ \\
 & 100 & 283.1 & $time$ & 283.5 & $time$ & 283.7 & $time$ & & 283.0 & $time$ & 283.4 & $time$ & 283.8 & $time$ \\
 & 225 & 255.0 & $time$ & 255.1 & $time$ & 255.4 & $time$ & & 255.0 & $time$ & 255.1 & $time$ & 255.3 & $time$ \\
 \hline
\end{tabular}
\label{tab:LS_DH}
\end{table}

It is crucially important to locate the sinks closer to their assigned sensors to improve the network lifetime $L$. In such a case, one activates fewer sensors, i.e., consumes less energy, to provide communication among the sensors and their assigned sinks. As we can see from Table \ref{tab:LS_DH}, increasing the budget level extends the lifetime $L$ at a certain energy level. For example, at low energy level and $(S = 2, N = 81)$, the lifetime $L$ gradually takes values 83.0, 83.9 and 84.7 as the budget level gets higher. Moreover, increasing the number of sinks in the network from $S = 2$ to 3 helps to the network lifetime $L$.

\begin{table}[h]
\centering
\tiny
\caption{Computational results for TS with DH}
\begin{tabular}{ccccccccccccccc} 
\hline
& & \multicolumn{6}{c}{$S = 2$} & & \multicolumn{6}{c}{$S = 3$} \\ \cline{3 - 8} \cline{10 - 15} \vspace{1mm}
& Budget & \multicolumn{2}{c}{Low} & \multicolumn{2}{c}{Medium} & \multicolumn{2}{c}{High} & &  \multicolumn{2}{c}{Low} & \multicolumn{2}{c}{Medium} & \multicolumn{2}{c}{High} \\ \cline{3 - 8} \cline{10 - 15}
\multirow{2}{*}{Energy} & \multirow{2}{*}{$N$} & \multirow{2}{*}{$L$} & CPU & \multirow{2}{*}{$L$}  & CPU &  \multirow{2}{*}{$L$} & CPU & &  \multirow{2}{*}{$L$} & CPU &  \multirow{2}{*}{$L$}  & CPU &  \multirow{2}{*}{$L$}  & CPU \\
& & & (secs) & & (secs)  & &  (secs) & & & (secs) & &  (secs)  & &  (secs) \\ 
 \hline
\multirow{8}{*}{Low} & 16 & 84.0 & 5.26 & 84.1 & 5.21 & 84.1 & 5.21 & & 84.1 & 16.25 & 84.1 & 16.25 & 84.2 & 16.23 \\
 & 25 & 87.5 & 28.91 & 87.6 & 28.73 & 87.8 & 29.72 & & 87.8 & 213.77 & 87.9 & 221.80 & 88.1 & 208.79 \\
 & 36 & 85.4 & 132.69 & 85.7 & 135.51 & 85.7 & 131.64 & &  85.3 & 1513.74 & 85.4 & 1548.78 & 85.7 & 1567.73 \\
 & 49 & 96.2 & 1675.13 & 96.3 & 1689.34 & 96.8 & 1642.58 & & 96.2 & $time$ & 96.4 & $time$ & 96.9 & $time$ \\
 & 64 & 79.2 & 3295.36 & 79.6 & 3314.54 & 80.7 & 3475.59 & & 79.3 & $time$ & 80.1 & $time$ & 81.0 & $time$ \\
 & 81 & 83.0 & $time$ & 83.9 & $time$ & 84.7 & $time$ & & 83.0 & $time$ & 83.7 & $time$ & 84.6 & $time$ \\
 & 100 & 90.8 & $time$ & 91.2 & $time$ & 91.5 & $time$ & & 90.6 & $time$ & 91.0 & $time$ & 91.4 & $time$ \\
 & 225 & 75.6 & $time$ & 75.7 & $time$ & 76.0 & $time$ & &  75.6 & $time$ & 75.7 & $time$ & 75.9 & $time$ \\ \hline
\multirow{8}{*}{Medium} & 16 & 167.6 & $time$ & 167.7 & $time$ & 167.7 & $time$ & & 167.7 & $time$ & 167.7 & $time$ & 167.8 & $time$ \\
 & 25 & 174.9 & $time$ & 175.0 & $time$ & 175.2 & $time$ & & 175.2 & $time$ & 175.3 & $time$ & 175.5 & $time$ \\
 & 36 & 175.1 & $time$ & 175.1 & $time$ & 175.1 & $time$ & & 174.9 & $time$ & 175.0 & $time$ & 175.1 & $time$ \\
 & 49 & 197.6 & $time$ & 197.7 & $time$ & 198.1 & $time$ & & 197.6 & $time$ & 197.8 & $time$ & 198.0 & $time$ \\
 & 64 & 165.0 & $time$ & 165.4 & $time$ & 166.3 & $time$ & & 165.1 & $time$ & 165.9 & $time$ & 166.8 & $time$ \\
 & 81 & 172.7 & $time$ & 173.6 & $time$ & 174.1 & $time$ & & 172.7 & $time$ & 173.3 & $time$ & 174.1 & $time$ \\
 & 100 & 185.8 & $time$ & 186.2 & $time$ & 186.4 & $time$ & & 185.6 & $time$ & 186.0 & $time$ & 186.4 & $time$ \\
 & 225 & 165.3 & $time$ & 165.4 & $time$ & 165.7 & $time$ & & 165.3 & $time$ & 165.4 & $time$ & 165.6 & $time$ \\ \hline
\multirow{8}{*}{High} & 16 & 253.4 & $time$ & 253.5 & $time$ & 253.5 & $time$ & & 253.5 & $time$ & 253.5 & $time$ & 253.6 & $time$ \\
 & 25 & 264.6 & $time$ & 264.7 & $time$ & 264.9 & $time$ & & 264.9 & $time$ & 265.0 & $time$ & 265.2 & $time$ \\
 & 36 & 264.8 & $time$ & 264.8 & $time$ & 264.8 & $time$ & & 264.6 & $time$ & 264.7 & $time$ & 264.8 & $time$ \\
 & 49 & 296.4 & $time$ & 296.5 & $time$ & 296.8 & $time$ & & 296.4 & $time$ & 296.6 & $time$ & 296.8 & $time$ \\
 & 64 & 250.8 & $time$ & 251.2 & $time$ & 251.5 & $time$ & & 250.9 & $time$ & 251.3 & $time$ & 251.8 & $time$ \\
 & 81 & 260.1 & $time$ & 260.9 & $time$ & 261.5 & $time$ & & 260.1 & $time$ & 260.7 & $time$ & 261.5 & $time$ \\
 & 100 & 283.1 & $time$ & 283.5 & $time$ & 283.7 & $time$ & & 283.0 & $time$ & 283.4 & $time$ & 283.8 & $time$ \\
 & 225 & 255.0 & $time$ & 255.1 & $time$ & 255.4 & $time$ & & 255.0 & $time$ & 255.1 & $time$ & 255.3 & $time$ \\
 \hline
\end{tabular}
\label{tab:TS_DH}
\end{table}

We summarize our results for TS method in Table \ref{tab:TS_DH}. We observe that TS takes longer time but provides better $L$ values compared to LS method. This is since TS searches larger portion of the $s-$swap neighborhood than LS (see $P_s(\%)$ values in Table \ref{tab:locp}). As an example, at the  low budget level with $(S = 2, N = 49)$, TS extends the lifetime $L$ of LS for all energy levels. In particular, for the high energy level $L=273.6$ for LS, whereas it is 296.4 for TS. Similar to LS, deploying more sinks elevates the lifetime $L$.

\section{Conclusions} \label{Conclusions}

We consider the sink location problem (SLP), connected coverage problem (CCP), activity scheduling problem (ASP), sink assignment problem (SAP) and data routing problem (DRP) to design heterogeneous WSNs. We propose a mixed integer programming formulation, i.e., $SPSRC$, that combines all design issues in a single model. $SPSRC$ finds the optimal locations of the sensors and sinks, active/standby periods of the sensors and the data transmission routes from each active sensor to its assigned sink. At each period, we need to cover each node in the network and the active sensors should communicate with each other to reach their assigned sinks.  The aim is to maximize the number of such periods within limited budget and battery energy resources.

$SPSRC$ is NP--complete since it includes the set covering problem as a subproblem. Hence, exact solution of the problem cannot be found even for small networks in acceptable amount of time.  For the solution of the problem, we first assume that the sink locations are given. For the reduced problem, i.e., $PSRC$, we propose constructive heuristic (CH) and disjunctive heuristic (DH). Then, we develop local search (LS) and tabu search (TS) methods to determine the best locations of the sinks for the network lifetime. 
In our computational experiments, we observe that DH is better than CH in terms of lifetime. Hence, we proceed with DH while experimenting for LS and TS. TS scans larger proportion of the solution space compared to LS. That is TS provides higher lifetime values within the time limit. 
Our solution techniques are heuristic approaches. That is, development of optimization algorithms for $SPSRC$ problem can be a future research. Furthermore, we assume that the sinks are at fixed locations. The problem can be further investigated for mobile sinks. 




\begingroup
    \setstretch{1.45}
\bibliographystyle{apa}
\bibliography{bibliography}
\endgroup

\end{document}